\documentclass[aps,pra,reprint,floatfix]{revtex4-1}
\usepackage{graphicx}
\usepackage{verbatim}
\usepackage[breaklinks=true,colorlinks,citecolor=blue,linkcolor=blue,urlcolor=blue]{hyperref}
\usepackage{amsmath}
\usepackage{xcolor}
\usepackage{footnote}
\DeclareMathOperator\erf{erf}
\begin{document}

\title{Universal nature of different methods of obtaining the exact Kohn-Sham exchange-correlation potential for a given density}

\author{Ashish Kumar}
\email[]{ashishkr@iitk.ac.in}
\affiliation{Department of Physics, Indian Institute of Technology Kanpur, Kanpur-208016, India}

\author{Rabeet Singh}
\email[]{rabeet@iitk.ac.in}
\affiliation{Department of Physics, Indian Institute of Technology Kanpur, Kanpur-208016, India}

\author{Manoj K. Harbola}
\email[]{mkh@iitk.ac.in\\
	   Presented in APS March Meeting 2018,\\ url-https://meetings.aps.org/Meeting/MAR18/Session/L60.325}
\affiliation{Department of Physics, Indian Institute of Technology Kanpur, Kanpur-208016, India}



\date{\today}

\begin{abstract}
 An interesting fundamental problem in density-functional theory  of electronic structure of matter is to construct the exact Kohn-Sham (KS) potential for a given  density. The exact potential can then be used to assess the accuracy of approximate functionals and the corresponding  potentials. Besides its practical usefulness, such a construction by itself is a challenging inverse problem. Over the past three decades, many seemingly disjoint methods have been proposed to  solve this problem. We show that these emanate from a single algorithm based on the Euler equation for the density. This provides a mathematical foundation for all different density-based  methods  that  are  used to construct the KS system from a given density and reveals their universal character.
\end{abstract}


\maketitle
Density functional theory (DFT) \cite{Yang, Drei, Gross,Hohenberg_PR.136.B864} is the most widely used method to study electronic properties of materials \cite{JoneRMP_89,SpruchRMP_1991, JoneRMP} because of its ever increasing accuracy \cite{SCANACC} and computational ease of implementation. As is well known, in DFT the ground-state energy is written as a functional $E[\rho]$ of the ground-state density $\rho(\vec{r})$. The energy is the sum of the kinetic energy functional $T[\rho]$, external energy $\int v_{ext}(\vec{r})\rho(\vec{r})d\vec{r}$ where $v_{ext}(\vec{r})$ is the external potential, and the expectation value  $\langle V_{ee}\rangle$, also a functional of $\rho(\vec{r})$, where ${V_{ee}}$ is the electron-electron interaction energy operator. In the Kohn-Sham \cite{Kohn_PR.140.A1133} approach to DFT (KSDFT) the interacting electron system is mapped to a fictitious non-interacting system and the energy $E[\rho]$ is expressed as the sum of the non-interacting kinetic energy $T_S[\rho]$ of the same density, the external energy, the Hartree energy $E_{H}[\rho]=\frac{1}{2}\iint\frac{\rho(\vec{r})\rho(\vec{r}')}{|\vec{r}-\vec{r}'|}d\vec{r}d\vec{r}'$ and the exchange-correlation energy $E_{xc}[\rho]$ with the difference $T_c= T[\rho]-T_{S}[\rho]$ absorbed in it. The equation for the density obtained by minimizing the energy $E[\rho]$ with respect to density is
\begin{equation}
\Big[ \frac{\delta T_S[\rho]}{\delta \rho}+ v_{ext}(\vec{r})+v_H(\vec{r})+v_{xc}(\vec{r})\Big]=\mu \label{euler},
\end{equation}
where $\mu$ is the Lagrange multiplier to ensure the constraint that $\int \rho(\vec{r})d\vec{r}=N=$ total number of electron and has the interpretation \cite{PPLB} of being the chemical potential. In Eq. \ref{euler}, the Hartree potential
\begin{equation}
v_H(\vec{r})=\int\frac{\rho(\vec{r}')}{|\vec{r}-\vec{r}'|}d\vec{r}' \label{hr_pot}
\end{equation}  
and the exchange-correlation potential 
\begin{equation}
v_{xc}(\vec{r})= \frac{\delta E_{xc}[\rho]}{\delta \rho(\vec{r})} \label{xc_pot} .
\end{equation}
We note that the exact expressions for $T_S[\rho]$ and $E_{xc}[\rho]$ in terms of the density are not known in general. Thus $\frac{\delta T_S[\rho]}{\delta \rho}$ and $v_{xc}[\vec{r}; \rho]$ are also not known exactly. In the Kohn-Sham(KS) formulation, the density is expressed in terms of single particle orbitals $\phi_i (\vec{r})$, occupying appropriately those with lowest energies so that $\rho(\vec{r})=\sum\limits_{i} f_i|\phi_i(\vec{r})|^2$ where $f_i$ is the occupancy of each orbital.  These orbitals are the solutions of the KS equation
\begin{equation}
\Big[ -\frac{1}{2}\nabla^2 +v_{ext}(\vec{r})+v_H(\vec{r})+v_{xc}(\vec{r})\Big]\phi_i(\vec{r})= \epsilon_i\phi_i(\vec{r}) \label{KS_eqn}.
\end{equation}
Although writen in terms of orbitals, Eq. \ref{KS_eqn} is equivalent to Eq. \ref{euler} for the density. Furthermore, the non-interacting kinetic energy $T_S[\rho]$ in the terms of KS orbitals is given exactly as
\begin{equation}
T_S[\rho]=\sum\limits_{i}f_i\langle\phi_i(\vec{r})|-\frac{1}{2}\nabla^2|\phi_i(\vec{r})\rangle \label{ks_ke} .
\end{equation}
Finally the orbital energies $\epsilon_i$ in general carry no physical meaning except that for the highest occupied \cite{PPLB} orbital which is equal to the chemical potential $\mu$.

\par In KSDFT, the  energy functional $E_{xc}[\rho]$ is approximated and consequently the KS equation can be solved only approximately. Development of more and more accurate functionals for  $E_{xc}[\rho]$ \cite{Medv_2017,TruhF} is of central interest for application of DFT since the accuracy of the energy and density obtained depends on the quality of $E_{xc}[\rho]$ and the corresponding $v_{xc}([\rho];\vec{r})$. 

\par While accurate exchange-correlation energy functionals are being developed \cite{Perdew_PRL.77.3865,Perdew_PRL.82.2544,Jianmin_PRL.91.146401,Sun_PNAS,Sun_PRL.115.036402,Becke_JCP.98.5648,Lee_PRB.37.785,Vosko_CJP.58.1200,Stephens_JPC.98.11623} and applied \cite{Chen_2017}, it is equally important to know the exact KS solution for a many-electron density wherever the latter is available. This gives \cite{Stott_1988,Gorling_1992,Zhao_1992, Zhao_1993,Wang_1993, Zhao_1994,Vlb_1994, WY1, WY2,Peirs_2003,Stott_2004, Viktor_2012, Viktor_2013, Wagner_2014, Viktor_2015,Wasserman_2017} the exact Kohn-Sham orbitals, the corresponding non-interacting kinetic energy and the related $T_c$. Furthermore, through construction of the KS systems, we also learn \cite{Buijse_1989,Gritsenko_1996,Teal2,Teal3,Teal4,Makmal_2011,Wagner_2012,Kohut_2016,Proetto_2016,Hollins_2017} about their other interesting aspects.  Thus, the exact KS systems set a benchmark to test the accuracy of approximate energy functionals.  We point out that for the exact density of electrons in a given external potential, constructing the KS system boils down to finding the exact exchange-correlation potential.

\par The problem of finding the KS system for a given ground-state density falls in the general category of inverse problems in physics \cite{Newton_1970}.  For a pedagogical review of such problems, we refer the reader to \cite{Carter_2000}.  In the present context, the direct problem is to find the ground-state density of a system of electrons in an external potential by solving the Schr\"odinger equation for the wavefunction.  The inverse problem \cite{Wasserman_2017}, whose solution is  warranted by the Hohenberg-Kohn theorem \cite{Hohenberg_PR.136.B864}, is to find the external potential or the wavefunction for a ground-state electronic density.  An interesting application \cite{Jayatilaka_PRL.80.798} of the inverse problem in this context has been to find the Hartree-Fock wavefunction for electrons in Beryllium crystal from its X-ray diffraction data. Finding the KS system for a ground-state density also falls in the same class of inverse problems and is of significant value in density-functional theory, as discussed above.

\par  Given its importance, many different methods have been developed over the years to obtain the exact Kohn-Sham system for a given density. Some of these \cite{WY1, WY2} are based on direct optimization of a functional while others are iterative \cite{Stott_1988,Gorling_1992,Zhao_1992, Zhao_1993, Zhao_1994, Wang_1993, Vlb_1994, Peirs_2003,Stott_2004, Wagner_2014}.  The latter methods converge towards the exact Kohn-Sham potential using a density based quantity to update the potential in each step of the iterative process. For example in reference \cite{Stott_2004}, the iterative method utilizes the difference between a given density and densities obtained during iterative steps to modify the potential. Interestingly, in the same paper iterative scheme has also been linked to an optimization method. However, in general a connection between different iterative methods and their relationship with the variational principle is not known.

\par The purpose of this paper is to show that all the inversion schemes (except that of \cite{Stott_1988}) referenced above are a result of obtaining the Levy-Lieb functional \cite{Levy_1979, Lieb_1983} for a given density and emanate from a single method that utilizes  Eq. \ref{euler} and  Eq. \ref{KS_eqn} in tandem. This method has its origin in the Levy-Perdew-Sahni (LPS) equation \cite{LPS_1984} for the density. Hence in the next section we first derive the method for the LPS equation and then generalize it to show how apparently different methods emerge from it. The general method is demonstrated by applying it in its different forms to some spherical system in section \ref{result}. Sections \ref{Gks} and \ref{result} thus reveal the universal character of all these methods. Using this universality, in section \ref{Theory} we prove that the inversion from density to Kohn-Sham system through any of these methods is equivalent to maximization of the functional $E[v]-\int v(\vec{r})\rho_0(\vec{r})d\vec{r}$ with respect to $v(\vec{r})$ to obtain the Levy-Lieb functional \cite{Levy_1979, Lieb_1983} for a given density $\rho_0(\vec{r})$; here $E[v]$ denotes the energy of the given number of electrons moving in the potential $v(\vec{r})$.  In the process we also derive a criterion for the convergence of the inversion process.  In section \ref{Equiv} we show  the equivalence of the general algorithm to different methods referenced above. Finally we conclude the paper in section \ref{summary}. 

\section{A general method to obtain the Kohn-Sham potential}
\label{Gks}
\subsection{Kohn-Sham potential from the LPS equation}
\label{GksA}
Consider the LPS equation for the density 
\begin{equation}
\big[-\frac{1}{2}\nabla^2 + v_{eff}(\vec{r})\big] \rho^{1/2}(\vec{r})= \mu\rho^{1/2}(\vec{r})\label{lps_eqn},
\end{equation}
where $v_{eff}(\vec{r})$  is given in the terms of the wavefunction. However, by writing the non-interacting kinetic energy as
\begin{equation}
T_S[\rho]=T_W[\rho] +T_P[\rho],
\end{equation}
where
\begin{equation}
T_W[\rho]=-\frac{1}{2}\int \rho^{1/2}(\vec{r}) \nabla^2\rho^{1/2}(\vec{r}) d\vec{r} \label{ws}
\end{equation}
is the Weizs\"acker kinetic energy or kinetic energy of Bosons of density $\rho(\vec{r})$ in the ground state, and $T_{P}[\rho]$ is the Pauli kinetic energy, it is easy to see that
\begin{equation}
v_{eff}(\vec{r})= v_{ext}(\vec{r})+v_P(\vec{r})+v_H(\vec{r})+v_{xc}(\vec{r}),
\end{equation} 
where $v_P=\frac{\delta T_P}{\delta \rho}$ is the Pauli potential \cite{MARCH_1985,LEVY_1988} . Thus with $\frac{\delta T_W}{\delta \rho}= -\frac{1}{2}\frac{\nabla^2\rho^{1/2}(\vec{r})}{\rho^{1/2}(\vec{r})}$ Eq. \ref{euler} for the density is
\begin{equation}
-\frac{1}{2}\frac{\nabla^2\rho^{1/2}(\vec{r})}{\rho^{1/2}(\vec{r})}+ v_{ext}(\vec{r})+v_P(\vec{r})+v_H(\vec{r})+v_{xc}(\vec{r})=\mu \label{LPS_KS}
\end{equation}
Now for a given exact density $\rho_0(\vec{r})$, if we denote the corresponding quantities with superscript  $\lq 0$', Eq. \ref{LPS_KS} can be rewritten for the exact exchange-correlation potential as
\begin{equation}
v_{xc}^{0}(\vec{r})= \mu+\frac{1}{2}\frac{\nabla^2\rho^{1/2}_0(\vec{r})}{\rho^{1/2}_0(\vec{r})}-v_{ext}(\vec{r})-v_P^0(\vec{r})-v_H^0(\vec{r}) \label{a}
\end{equation}
and for the exact Pauli potential as
\begin{equation}
v_{P}^{0}(\vec{r})= \mu+\frac{1}{2}\frac{\nabla^2\rho^{1/2}_0(\vec{r})}{\rho^{1/2}_0(\vec{r})}-v_{ext}(\vec{r})-v_H^0(\vec{r})-v_{xc}^0(\vec{r}) \label{b}
\end{equation}
Note that $\lq \mu$' is given by density $\rho_0(\vec{r})$ from its asymptotic behavior. We use Eq. \ref{a} and Eq. \ref{b} to write exchange-correlation potential corresponding to density $\rho_0(\vec{r})$ for $(i+1)^{th}$ iteration if at $i^{th}$ iteration the density is $\rho_i(\vec{r})$. Accordingly
\begin{equation}
v_{xc}^{i+1}(\vec{r})= \mu+\frac{1}{2}\frac{\nabla^2\rho^{1/2}_0(\vec{r})}{\rho^{1/2}_0(\vec{r})}-v_{ext}(\vec{r})-v_P^i(\vec{r})-v_H^i(\vec{r}), \label{c}
\end{equation}
where
\begin{equation}
v_{P}^i(\vec{r})= \mu^i+\frac{1}{2}\frac{\nabla^2\rho^{1/2}_i(\vec{r})}{\rho^{1/2}_i(\vec{r})}-v_{ext}(\vec{r})-v_H^i(\vec{r})-v_{xc}^i(\vec{r}). \label{d}
\end{equation}
Substituting Eq. \ref{d} in Eq. \ref{c} gives (We have dropped the constant term $\mu-\mu^i$.)
\begin{equation}
v_{xc}^{i+1}(\vec{r})=v_{xc}^i(\vec{r}) -\frac{1}{2}\frac{\nabla^2\rho^{1/2}_i(\vec{r})}{\rho^{1/2}_i(\vec{r})}+\frac{1}{2}\frac{\nabla^2\rho^{1/2}_0(\vec{r})}{\rho^{1/2}_0(\vec{r})}  \label{master1}
\end{equation}
The constant $(\mu-\mu^i)$ can be fixed either by adjusting the potential to get the correct $\mu$ or fixing its value at a large distance. Eq. \ref{master1} is the working equation for obtaining the exchange-correlation potential $v^0_{xc}(\vec{r})$ up to a constant for the ground state density $\rho_0(\vec{r})$.

\par Using Eq. \ref{master1}, the algorithm to find the exchange-correlation potential for a density $\rho_0(\vec{r})$ is as follows:
\begin{itemize}
	\item Start with a trial exchange-correlation potential $v_{xc}(\vec{r})$ and solve the KS equation to obtain the corresponding Kohn-Sham orbitals, the density and $\mu^i=\epsilon^{max}$. The external and Hartree potentials in the KS equations are the exact ones with the latter being calculated from the density $\rho_0({\vec{r}})$. 
	\par At the $i^{th}$ iteration this step gives the ground state density $\rho_i(\vec{r})$ corresponding to the exchange-correlation potential $v_{xc}^i(\vec{r})$. The density $\rho_i(\vec{r})$ also serves as an approximation to $\rho_0(\vec{r})$ and is expected to get closer to it with the increasing number of iterations ; 
	\item Find the new potential using Eq. \ref{master1}. At this step one can either use $\mu-\epsilon^{max}$ explicitly or fix the potential asymptotically by using the boundary condition for it;
	\item Use the new potential in the KS equation again until the density obtained from its solutions matches with the given density.
\end{itemize}
\par 
For completeness we point out that the expression for $v_{eff}(\vec{r})$ of Eq. \ref{lps_eqn} in terms of the many-body wavefunction was given by LPS \cite{LPS_1984} and has been employed  \cite{Buijse_1989,Gritsenko_1994,Gritsenko_1996,Gritsenko_1998} extensively to study properties of the  Kohn-Sham potential. Secondly, if $\frac{1}{2}\frac{\nabla^2 \rho^{1/2}_0}{\rho^{1/2}_0}$ and $\frac{1}{2}\frac{\nabla^2 \rho^{1/2}_i}{\rho^{1/2}_i}$ in Eq. \ref{master1} are replaced by $v_{eff} (\vec{r})$ derived from the true wavefunction and the Kohn-Sham orbitals in the  $i^{th}$ iteration, respectively, an expression for $v_{xc}^{i+1}$ is obtained  in terms of quantities that depend explicitly on the wavefunction and the KS orbitals. This approach has been utilized to get $v_{xc}$, or $v_{x}$ in Hartree-Fock (HF) theory, directly from wavefunctions and is discussed in the Appendix. 

\subsection{Use of a general functional \texorpdfstring{$S[\rho]$}{Lg} to obtain the KS potential} \label{GksB}
To generalize Eq. \ref{master1} to find the KS potential we split the kinetic energy functional $T_S[\rho]$ as
\begin{equation}
T_S[\rho]= S[\rho]+\tilde{T}_P[{\rho}],
\end{equation}
where $S[\rho]$ is a functional with the dimensions of energy and $\tilde{T}_P[{\rho}]= T_S[\rho]-S[\rho]$ is the generalized Pauli kinetic energy.  An important property of $S[\rho]$ will be derived in section (\ref{Theory}). In terms of $S[\rho]$, the equation for the density is 
\begin{equation}
\frac{\delta S}{\delta \rho}+v_{ext}(\vec{r})+\tilde{v}_P(\vec{r})+v_H[\rho(\vec{r})]+v_{xc}(\vec{r})= \mu, \label{gn-ks-lps}
\end{equation}
where $\tilde{v}_P =\frac{\delta \tilde{T}_P}{\delta \rho}$. Analogous to the manner in which Eq. \ref{LPS_KS} leads to Eq. \ref{master1} relating $v_{xc}^{i+1}(\vec{r})$ to $v_{xc}^{i}(\vec{r})$, Eq. \ref{gn-ks-lps} gives 
\begin{equation}
v_{xc}^{i+1}(\vec{r}) = v_{xc}^i(\vec{r})+\frac{\delta S}{\delta \rho}\Big|_{\rho _i(\vec{r})}-\frac{\delta S}{\delta \rho}\Big|_{\rho _0(\vec{r})}, \label{master2}
\end{equation}
where $\frac{\delta S}{\delta \rho}\Big|_{\rho_i(\vec{r})}$ implies that the functional derivative is evaluated at density $\rho_i(\vec{r})$. This is the general equation for obtaining the exchange-correlation potential $v_{xc}^0(\vec{r})$ corresponding to given density $\rho_0(\vec{r})$. Following the steps given at the end of section \ref{GksA}, it can be employed iteratively to obtain the exact exchange-correlation potential for a given density $\rho_0({\vec{r}})$ with the functional $S[\rho]$ of one's choice.  

\par Notice that if $S[\rho]$ is taken to be the Weizs\"acker functional $T_W[\rho]$, Eq. \ref{master1} is recovered.  However, with Eq. \ref{master2} we have the flexibility of choosing $S[\rho]$ to be a more general functional. For example, $S[\rho]$ can be chosen to be $\int f(\vec{r})\rho^n(\vec{r})d\vec{r}$ $(n>1) $, where $f(\vec{r})$ is an appropriately chosen function, or the Hartree energy $\frac{1}{2}\iint\frac{\rho(\vec{r})\rho(\vec{r}')}{|\vec{r}-\vec{r}'|}d\vec{r}d\vec{r}'  $.  This is somewhat along the lines of deriving the generalized density functional theory \cite{Siedl_1996} where the functional $\langle T+V_{ee}\rangle$ (or $  F[\rho]$) is split differently from KSDFT into the Hartree-Fock energy functional and a correlation energy functional.
\begin{figure*}
	\begin{center}
		\includegraphics[scale=0.85]{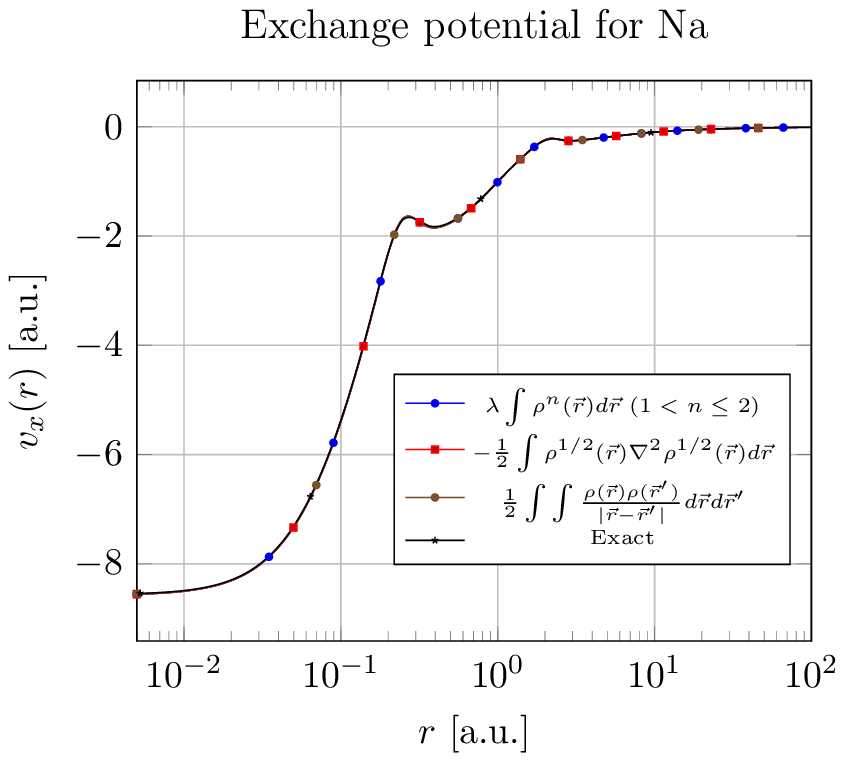} \hspace{0.5cm}
		\includegraphics[scale=0.85]{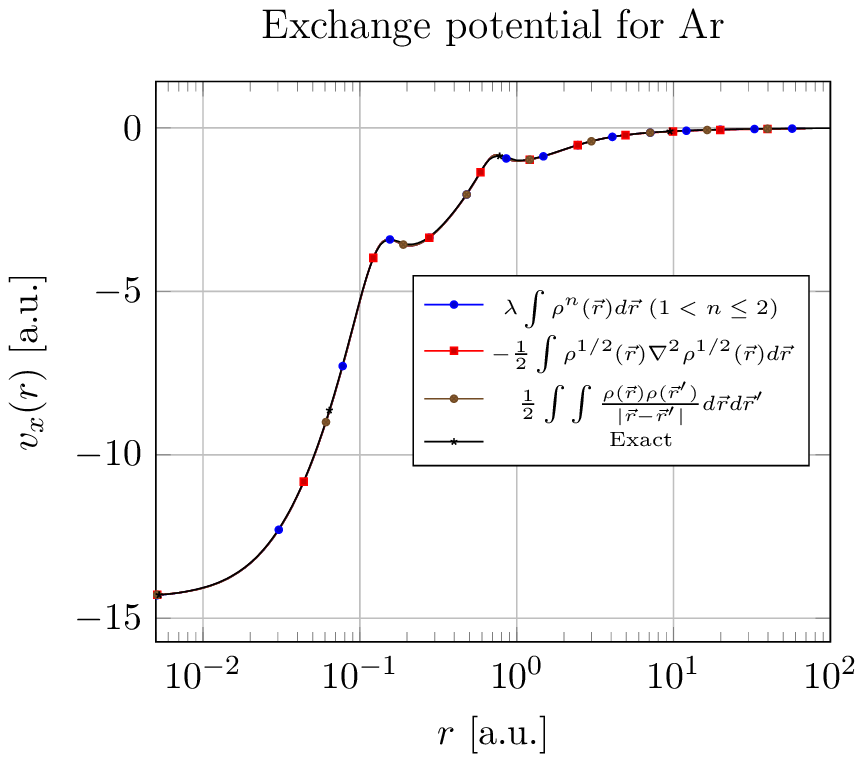} 
		\caption{ \label{Fig1} Exchange potential calculated for Na and Ar atom HF density using different form of $S[\rho]$ shown in the inset. In the functional $\lambda \int \rho^n (\vec{r}) d\vec{r}$, we have taken $\lambda=1$ and several values of $n$ between $1$ and $2$.}
	\end{center}
\end{figure*} 

\begin{figure} 	
	\begin{center}
		\includegraphics[scale=0.80]{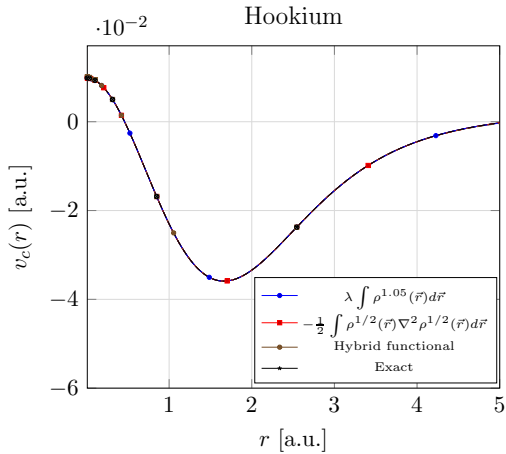}
		\caption{ \label{Fig2} Correlation potential calculated Hookium atom  using using different form of $S[\rho]$ in the inset.}
	\end{center}
\end{figure}	  
\begin{figure}
	\begin{center}
		\includegraphics[scale=0.80]{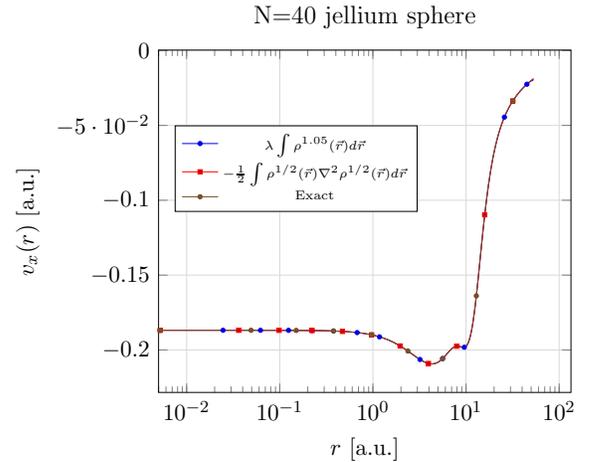}
		\caption{ \label{Fig3} Exchange potential calculated for N=40 jellium sphere from the Harbola-Sahni density. Different $S[\rho]$ used are shown in the inset.}
	\end{center} 	
\end{figure}
\subsection{Hybrid type functional \texorpdfstring{$S[\rho]$}{Lg}}
An advantage of having many functionals $S[\rho]$ that can be used in Eq. \ref{master2} is that we can choose different functionals in different regions of a system. This is useful if one particular functional $S_1[\rho]$ gives better convergence in one region of the system but some other functional $S_2[\rho]$ in other regions. For example, in the asymptotic regions where density is very small, the functional $S_1[\rho]=\int\rho^n(\vec{r})d\vec{r}$ with $n$ slightly larger than $1$ (for example $1.05$) gives accurate answers because $ \rho^{0.05}$ is relatively larger there. Thus, for spherical systems we can choose $S[\rho]$ so that 
\begin{eqnarray}
\frac{\delta S[\rho]}{\delta \rho}= \erf(\alpha r)\rho^{0.05} +  (1-\erf(\alpha r))\rho^{0.5}, \label{hyb_eq}
\end{eqnarray} 
where $\erf(\alpha r)$ is the error function with a suitably chosen parameter $\alpha$. We make use of such mixing of different $S[\rho]$ in the results section below.

\section{Results} \label{result}
We now demonstrate the ideas presented above through spherically symmetric systems. All the numerical calculation of these systems are carried out using the Herman-Skillman program \cite{Herman_PHP} by modifying it suitably.   
In Fig. \ref{Fig1} we show the exchange potential for the  Hartree-Fock density of Na and Ar 
atoms \cite{Bunge_1993} using the functionals: 
\[
S[\rho] = 
\begin{cases}
-\frac{1}{2} \int \rho(\vec{r})^{1/2}\nabla^2\rho(\vec{r})^{1/2}d\vec{r}               & (i) \\
\\
\int\rho^n(\vec{r})d\vec{r}, \quad \quad (1<n\leq 2)    &   (ii)\\ \\
\frac{1}{2}\iint\frac{\rho(\vec{r})\rho(\vec{r}')}{|\vec{r}-\vec{r}'|}d\vec{r}d\vec{r}'& (iii)
\end{cases}
\]
The output exchange potential matches with the corresponding optimized effective potential (OEP) \cite{Talman_1978} for all the functional forms mentioned above with $n$ varying over a large number of values. We have also done calculation with $\frac{\delta S[\rho]}{\delta \rho}$ given in Eq. \ref{hyb_eq} and found that due to $\rho^{0.05}$ in it, the potential in the asymptotic region is reproduced with ease. We comment on this further in the paragraph below.

\par Next in Fig. \ref{Fig2} we display the correlation potential of Hookium atom calculated using $S[\rho] =-\frac{1}{2} \int \rho(\vec{r})^{1/2}\nabla^2\rho(\vec{r})^{1/2}d\vec{r}$, $\int \rho^{1.05}(\vec{r}) d\vec{r}$ and the hybrid functional Eq. \ref{hyb_eq}. The output potential matches perfectly with the exact correlation potential \cite{Kais_1993}. We wish to point out that although functional $S[\rho]= \int \rho^{1.05}(\vec{r}) d\vec{r}$ gives exact result here, for inner regions a functional with large power of $\rho$ is equally good. It is in the outer regions where the density becomes very small of because of its $e^{-r^2}$ dependence on $r$ that the functional $\int \rho^{1.05}(\vec{r}) d\vec{r}$ become really useful. This is a good example of how hybrid functionals are effective in such situations. 
\par Finally in Fig. \ref{Fig3}, we employ the density for a neutral jellium sphere \cite{Knight_PRL.52.2141,Matthias_RMP.65.677} to get the KS exchange potential. These densities are obtained using the Harbola-Sahni (HS) exchange potential \cite{MKH_89} and the general method of Eq. \ref{master2} reproduces it with two different forms of  $S[\rho]$ viz. $\int \rho^{1.05}(\vec{r}) d\vec{r}$ and $- \frac{1}{2} \int \rho(\vec{r})^{1/2}\nabla^2\rho(\vec{r})^{1/2}d\vec{r}$ .  For other forms stated above  deviation from the exact potential starts for $r>20$ as the density becomes very low.

\par We note that, to the  best of our knowledge, the Weizs\"acker functional has not been used in the past to get the Kohn-Sham potential. The functional $\int f(\vec{r}) \rho^n(\vec{r})d \vec{r}$  has been employed taking $n=2$, with $f(\vec{r}) =1$ \cite{Wasserman_2017} and $f(\vec{r}) = r^{\beta} (0<\beta<3)$ \cite{Peirs_2003}. Recently the Hartree functional has been applied \cite{Hollins_2017} to get the local exchange potential for Hartree-Fock density of solids.

\section{Theory}
\label{Theory}
The density $\rho_0(\vec{r})$ to potential $v(\vec{r})$ map can be established  through the Levy-Lieb functional \cite{Levy_1979, Lieb_1983}  that is defined by Lieb \cite{Lieb_1983} as
\begin{equation}
F[\rho_0]= \underset{v}{\text{Supremum}}\Big[ E[v]-\int v(\vec{r})\rho_0(\vec{r})d\vec{r}\Big] \label{lieb}
\end{equation}
where the search for the  Supremum is done over different potentials. For the true ground-state densities the Supremum is a maximum. We now prove that with properly chosen $S[\rho]$ the method of section \ref{GksB} (also therefore all such method that it encompasses) makes $E[v]-\int v(\vec{r})\rho_0(\vec{r})d\vec{r}$ larger and larger in each iterative step converging finally to the correct $F[\rho_0]$. To this end we consider the potentials $v^i$ and $v^{i+1}$ for the $i^{th}$ and $(i+1)^{th}$ steps and calculate the difference
\begin{eqnarray*}
	\Delta F &=& \Big( E[v^{i+1}]-\int v^{i+1}(\vec{r})\rho_0(\vec{r})d\vec{r}\Big)\\
	& -& \Big( E[v^i]-\int v^i(\vec{r})\rho_0(\vec{r})d\vec{r}\Big)\\
	& =&E[v^{i+1}]-E[v^{i}] -\int(v^{i+1}-v^i)\rho_0(\vec{r})d\vec{r}.
\end{eqnarray*} 
For small $(v^{i+1}-v^{i})$ - and this can always be ensured by proper mixing \cite{Wagner_2013} of $v^i$ and the potential calculated from equation (\ref{master2}) - we have by the first order perturbation theory
\begin{equation*}
E[v^{i+1}]-E[v^i]= \int (v^{i+1}(\vec{r})-v^i(\vec{r}))\rho_i(\vec{r})d\vec{r}
\end{equation*}
and therefore
\begin{eqnarray}
&\Delta F& = \int(v^{i+1}(\vec{r})-v^i(\vec{r}))(\rho_i(\vec{r})-\rho_0(\vec{r}))d\vec{r} \nonumber\\
&=& \int \Big(\frac{\delta S}{\delta \rho}\Big|_{\rho _i(\vec{r})}-\frac{\delta S}{\delta \rho}\Big|_{\rho _0(\vec{r})}\Big)(\rho_i(\vec{r})-\rho_0(\vec{r}))d\vec{r} \label{cond}.
\end{eqnarray}
If the iterative process is to converge towards the correct potential, from Eq. \ref{lieb} we should have $\Delta F \geq 0$ at each iterative step with the equality being satisfied when $\rho_i= \rho_0$. As indicated by Eq. \ref{cond}, this will be the case if the functional $S[\rho]$ is such that
\begin{equation}
\int \Big(\frac{\delta S}{\delta \rho}\Big|_{\rho _i(\vec{r})}-\frac{\delta S}{\delta \rho}\Big|_{\rho_0(\vec{r})}\Big)(\rho_i(\vec{r})-\rho_0(\vec{r}))d\vec{r} \geq 0 \label{cond1}.
\end{equation}
Eq. \ref{cond1} therefore is the condition on $S[\rho]$ for finding the Kohn-Sham potential using iterative methods. We note that for the KS system, $F[\rho_0]=T_S[\rho_0]$.  The proof above is akin to the demonstration \cite{Wagner_2013} that the iterative Kohn-Sham solution always converges towards minimum energy. A strong condition on $S[\rho]$ will be that the integrand  
\begin{equation}
\Big(\frac{\delta S}{\delta \rho}\Big|_{\rho _i(\vec{r})}-\frac{\delta S}{\delta \rho}\Big|_{\rho _0(\vec{r})}\Big)(\rho_i(\vec{r})-\rho_0(\vec{r})) \geq0 \label{cond2}.
\end{equation}
This condition is easy to understand physically: it means that in each iterative step the potential increases (decreases) if the density $\rho(\vec{r})$ in the previous step is larger(smaller) than the target density $\rho_0(\vec{r})$.

\par To sum up, we have shown that with a properly chosen $S[\rho]$, the process of obtaining the Kohn-Sham potential for a given density converges by maximizing $E[v]-\int v\rho_0(\vec{r})d\vec{r}$ iteratively. Therefore the iterative method is equivalent to the direct optimization method \cite{WY2,Stott_2004} for finding the Kohn-Sham system. This is complementary to the equivalence of the minimization of energy functional for a given $v_{ext}(\vec{r})$ and solving the corresponding Kohn-Sham equations to get the ground-state density \cite{Wagner_2013}.  It further requires the corresponding functional $S[\rho]$ to satisfy the condition given by Eq. \ref{cond1} or Eq. \ref{cond2}. 
\par It is easy to see that the convergence condition on  $S[\rho]$ is satisfied in its strong form Eq. \ref{cond2} for the functionals $S[\rho]= \int \rho^n(\vec{r})d\vec{r}$ with $n>1$. It is also satisfied for  $S[\rho] = \frac{1}{2}\iint\frac{\rho(\vec{r})\rho(\vec{r}')}{|\vec{r}-\vec{r}'|}d\vec{r}d\vec{r}'$ since in that case Eq. \ref{cond1} is equivalent to  $ \frac{1}{2}\iint\frac{(\rho(\vec{r})-\rho_0(\vec{r}))(\rho(\vec{r}')-\rho_0(\vec{r}')}{|\vec{r}-\vec{r}'|}d\vec{r}d\vec{r}' \geq 0 $ which is always true. We now show this to be true for the Weizs\"acker functional also. In that case the condition is 
\begin{equation*}
\int \Big( \frac{\nabla^2 \rho_0^{1/2}(\vec{r})}{\rho_0^{1/2}(\vec{r})} - \frac{\nabla^2 \rho^{1/2}(\vec{r})}{\rho^{1/2}(\vec{r})}\Big)(\rho(\vec{r})-\rho_0(\vec{r}))d\vec{r} \geq 0.
\end{equation*}
The condition can easily be shown to be equivalent to 
\begin{eqnarray*}
	& &	\int \Big[ \nabla \rho^{1/2}(\vec{r}) -\Big(\frac{\rho(\vec{r})}{\rho_0(\vec{r})}\Big)^{1/2}\nabla \rho_0^{1/2}(\vec{r})\Big]^2 d\vec{r} \\
	&+& \int \Big[ \nabla \rho_0^{1/2}(\vec{r}) -\Big(\frac{\rho_0(\vec{r})}{\rho(\vec{r})}\Big)^{1/2}\nabla \rho^{1/2}(\vec{r})\Big]^2 d\vec{r} \geq 0
\end{eqnarray*} 
which is always satisfied.

\par Finally we note that recently the method of obtaining a local potential for a given wavefunction \cite{Viktor_2013,Viktor_2015} discussed in the Appendix has also been related \cite {Gidopoulos_2011,Teal_2017} to finding the Levy-Lieb functional. In essence it is also equivalent to finding Levy-Lieb functional by maximizing $E[v]-\int v(\vec{r})\rho_0(\vec{r})d\vec{r}$.
\section{Equivalence of different iterative methods}
\label{Equiv}
In this section we show that different density-based inversion schemes suggested in the literature are equivalent to using an appropriate functional $S[\rho]$. We consider them one by one in the following:\\
\subsection{van-Leeuwen Baerends (vLB) method \texorpdfstring{\cite{Vlb_1994,Vlb_1995}}{Lg} and its variants \texorpdfstring{\cite{Wasserman_2017, Stott_2004}}{Lg}}
In the vLB method, the potential $v_{Hxc}(\vec{r})= v_H(\vec{r})+v_{xc}(\vec{r})$ due to the electron-electron interaction is updated in each cycle as
\begin{eqnarray}
v_{Hxc}^{i+1}(\vec{r})&=& \frac{\rho_{i}(\vec{r})}{\rho_0(\vec{r})}v_{Hxc}^i(\vec{r}) \notag \\
&=&\frac{(\rho_{i}(\vec{r})-\rho_0(\vec{r}))}{\rho_0(\vec{r})}v_{Hxc}^i(\vec{r})+v_{Hxc}^i(\vec{r}). \label{vlbeq}
\end{eqnarray}
Thus the functional 
\begin{equation*}
S[\rho] = \frac{1}{2} \int \frac{v^i_{Hxc}(\vec{r})}{\rho_0(\vec{r})}\rho^2(\vec{r})d\vec{r},
\end{equation*}
where $v^i_{Hxc}(\vec{r})$ is Hartree-exchange-correlation potential for the $i^{th}$ iteration, leads to Eq. \ref{vlbeq} when substituted in Eq. \ref{master2}. It is pointed out that $v^i_{Hxc}(\vec{r})$ is the potential in the $i^{th}$ iteration and therefore remains unchanged when $\rho$ is varied. Furthermore
\begin{equation}
\Delta F = \int \frac{v^i_{Hxc}(\rho(\vec{r})-\rho_0(\vec{r}))^2}{\rho_0(\vec{r})}d\vec{r} \ge 0
\end{equation}  
so that the procedure satisfies the condition for it to converge. Other variants of the method are with different powers of density in $S[\rho]$ or that given in \cite{Wasserman_2017}. Among these we note that the method of Wang and Parr \cite{Wang_1993} will converge only if bound states have negative eigenvalues. 

\par An alternative to the vLB method for $v_{ext}(\vec{r})<0$ is obtained with
\begin{equation*}
S[\rho] = - \frac{1}{2} \int \frac{v_{ext}(\vec{r})}{\rho_0(\vec{r})}\rho^2(\vec{r})d\vec{r},
\end{equation*}
so that 
\begin{equation}
v_{Hxc}^{i+1}(\vec{r})= -v_{ext}(\vec{r})\frac{(\rho_{i}(\vec{r})-\rho_0(\vec{r}))}{\rho_0(\vec{r})}+v_{Hxc}^i(\vec{r}) \label{vlbeq2}.
\end{equation}
The negative sign here is to ensure that for $v_{ext}(\vec{r})<0$ the convergence condition Eq. \ref{cond1} satisfied. For $v_{ext}(\vec{r})>0$, the sign above will be positive.
\subsection{G\"orling \texorpdfstring{\cite{Gorling_1992}}{Lg}, Gaudoin and Burke method \texorpdfstring{\cite{Wagner_2014,Gaudoin_2004}}{Lg}}
In this method, the change in potential is calculated using 
\begin{equation}
v^{i+1}(\vec{r})-v^i(\vec{r})= \int \chi^{-1}_i(\vec{r}, \vec{r}')(\rho_0(\vec{r}')-\rho(\vec{r}'))d\vec{r}' \label{buke},
\end{equation}
where $\chi^{-1}_i(\vec{r}, \vec{r}')[\rho_i]$ is the non-interacting inverse response function  for the system at $i^{th}$ iteration with exchange-correlation potential $v_{xc}^i(\vec{r})$ and density $\rho_i(\vec{r})$.
Thus if we take $S[\rho]= -\frac{1}{2}\iint \chi^{-1}_i(\vec{r},\vec{r}')\rho(\vec{r})\rho(\vec{r}')d\vec{r}'d\vec{r}$ we get the updated potential as given by Eq. \ref{buke} . Observe that while taking the functional derivative of $S[\rho]$, the inverse response function $\chi^{-1}_i(\vec{r},\vec{r}')$ does not contribute to it because it is independent of the variable $\rho(\vec{r})$. Therefore 
\begin{eqnarray*}
	\Delta F= -\iint \chi^{-1}_i(\vec{r},\vec{r}')(\rho_0(\vec{r}')-\rho(\vec{r}'))(\rho(\vec{r})-\rho_0(\vec{r}))d\vec{r}'d\vec{r}.
\end{eqnarray*} 
Following \cite{Vlb_2003}, it is easy to show that $$ \iint \chi^{-1}_i(\vec{r},\vec{r}') f(\vec{r})f(\vec{r}')d\vec{r}'d\vec{r} < 0$$ 
for a function $f(\vec{r})$.  The iterative scheme therefore follows the convergence criterion to maximize $E[v]-\int v(\vec{r})\rho_0(\vec{r}) d\vec{r}$ with respect to $v(\vec{r})$.

\subsection{Peirs, Van Neck and Waroquier (PNW)  method \texorpdfstring{\cite{Peirs_2003}}{Lg}}
PNW use the following update algorithm to find the exchange-correlation potential for spherical systems:
\begin{equation}
v_{xc}^{i+1}(r)=v_{xc}^i(r)+ \lambda r^{\beta} (\rho_i -\rho_0) + f(r)(\mu^i -\mu^0), \label{pnweq1}
\end{equation}
where $0.5 <\lambda <3.5$ and $0<\beta<3$. The function $f(r)$ is a switching function  used to tune the asymptotic behavior of potential. Leaving the last term in Eq. \ref{pnweq1}, the functional leading to the PNW algorithm is
\begin{equation}
S[\rho]=\lambda\int r^\beta \rho^n(\vec{r})d\vec{r} \label{pnweq2}
\end{equation}
with $n=2$ and an optimized value of $\lambda$, $\beta$. It is easy to see that functional satisfies the strong condition of Eq. \ref{cond2} for the convergence of the algorithm.
\par The present work implies that the PVN method can be generalized by using any $n>1$ in Eq. \ref{pnweq2}.
\subsection{Hollins, Clark, Refson and Gidopoulos (HCRG)  method \texorpdfstring{\cite{Hollins_2017}}{Lg}}
As pointed out earlier, recently  the Hartree potential has been used by HCRG \cite{Hollins_2017} to calculate the exchange-correlation potential corresponding to the Hartree-Fock density. In this method the exchange-correlation potential is updated according to the equation
\begin{equation}
v^{i+1}_{xc}(\vec{r}) = v^{i}_{xc}(\vec{r}) + \epsilon \int \frac{\rho_i(\vec{r}') -\rho_0(\vec{r}')}{|\vec{r}-\vec{r}'|}d\vec{r}' \label{hcrg_eqn};
\end{equation}	
where $\epsilon$ is small positive number.
As is evident, the functional $S[\rho]=  \frac{\epsilon}{2} \iint \frac{\rho(\vec{r})\rho(\vec{r}')}{|\vec{r}-\vec{r}'|}d\vec{r}'d\vec{r}$ gives the working  Eq. \ref{hcrg_eqn}. We have already discussed that this $S[\rho]$  satisfies the condition  for convergence given by Eq. \ref{cond1}.
\subsection{Zhao-Maorrison-Parr (ZMP)  \texorpdfstring{\cite{Zhao_1994}}{Lg} mehod}
In the ZMP method, the KS potential is obtained as the Hartree-potential of difference in the given density $\rho_0(\vec{r})$ and the solution density $\rho(\vec{r})$ multiplied by a large constant $\lambda$. The equation to be solved in the ZMP method is \cite{Zhao_1994}
\begin{equation}
\Big[ -\frac{1}{2}\nabla^2 +v_{ext}(\vec{r})+(1-\frac{1}{N})v_H(\vec{r}) + v_{ZMP}(\vec{r}) \Big]\phi_i= \epsilon_i \phi_i \label{zmp_eqn}
\end{equation}
with
\begin{equation}
v_{ZMP}(\vec{r}) =\lambda \int \frac{\rho(\vec{r}')-\rho_0(\vec{r}')}{|\vec{r}-\vec{r}'|}d\vec{r}'.
\end{equation}
Here $v_H(\vec{r})$ is Hartree potential of given density $\rho_0(\vec{r})$ and $\rho(\vec{r})= \sum_{i} |\phi_i(\vec{r})|^2$. We point out that the self-interaction component of the exchange-correlation potential has been included with Hartree potential and that make achieving self-consistency easier. In using this method, one usually starts with a small value of $\lambda$ and then increases it to obtain better and better density $\rho(\vec{r})$. Finally the exchange-correlation potential is obtained as
\begin{equation}
v_{xc}(\vec{r})= \lim\limits_{\lambda \to \infty}  v_{ZMP}(\vec{r})-\frac{v_H(\vec{r})}{N}.
\end{equation}
For a given $\lambda$ Eq. \ref{zmp_eqn} is solved self-consistently. Thus one starts with  some initial guess of $v_{ZMP}(\vec{r})$, say $v^1_{ZMP}(\vec{r})$, and at the $(i+1)^{th}$ cycle of the self-consistent procedure the potential is updated as 
\begin{eqnarray}
v^{i+1}_{ZMP}(\vec{r}) =(1-\alpha)v^i_{ZMP}(\vec{r})+ \alpha \lambda \int\frac{\rho_i(\vec{r}')-\rho_0(\vec{r}')}{|\vec{r}-\vec{r}'|}d\vec{r}' \label{zmp_eqn2},
\end{eqnarray}
where $\alpha (< 1)$ is the mixing parameter and $v^i_{ZMP}(\vec{r})$  is the  potential for the $i^{th}$ iteration with solution density $\rho_i(\vec{r})$. If one takes $v^1_{ZMP}(\vec{r}) =0$ then Eq. \ref{zmp_eqn2} leads to
\begin{eqnarray}
v^{i+1}_{ZMP}(\vec{r}) =\alpha \lambda\Big[\sum_{m=1}^{i} (1-\alpha)^{i-m}\int\frac{\rho_m(\vec{r}')-\rho_0(\vec{r}')}{|\vec{r}-\vec{r}'|}d\vec{r}' \label{zmp_eqn3}\Big].
\end{eqnarray}
In the HCRG method, the potential $v_{ZMP}(\vec{r})$ in Eq. \ref{zmp_eqn} is replaced by $v_{HCRG}(\vec{r})$ and it is updated as (by taking $\epsilon = \alpha$)
\begin{equation}
v^{i+1}_{HCRG}(\vec{r}) = v^i_{HCRG}(\vec{r})+ \alpha\int\frac{\rho_i(\vec{r}')-\rho_0(\vec{r}')}{|\vec{r}-\vec{r}'|}d\vec{r}'\label{hcrg_eqn2}.
\end{equation}
After achieving the convergence, the exchange-correlation potential $v_{xc}(\vec{r})$ is calculated as
\begin{equation}
v_{xc}(\vec{r})= v_{HCRG}(\vec{r})- \frac{v_H(\vec{r})}{N}.
\end{equation}
Again taking $v^1_{HCRG}(\vec{r}) =0$, Eq. \ref{hcrg_eqn2} becomes
\begin{eqnarray}
v^{i+1}_{HCRG}(\vec{r}) =\alpha\Big[\sum_{m=1}^{i} \int\frac{\rho_m(\vec{r}')-\rho_0(\vec{r}')}{|\vec{r}-\vec{r}'|}d\vec{r}' \label{hcrg_eqn3}\Big].
\end{eqnarray}
From Eq. \ref{zmp_eqn3} and Eq. \ref{hcrg_eqn3} it is evident that the methods of ZMP and HCRG are equivalent as both  use the electrostatic potential of charge density $\rho_i(\vec{r})-\rho_0(\vec{r})$ for improvement of the potentials at each iterative step. However, the way these corrections are added during the process is different. In the ZMP method (Eq. \ref{zmp_eqn3}), the contribution of the potentials from previous iterations keeps on diminishing as the number of iterations increases and self-consistency is approached with the density difference becoming smaller and smaller. Thus to keep the potential finite, a large value of $\lambda$ is needed. On the other hand, in the HCRG method (Eq. \ref{hcrg_eqn3}), potential at each iteration contributes equally.
\par We note that to satisfy the convergence condition of Eq. \ref{cond1} for the ZMP method, the value of $\alpha$ should be very small. This is because
\begin{eqnarray}
v^{i+1}_{ZMP}(\vec{r})-v^{i}_{ZMP}(\vec{r}) = \lambda \Big[  \alpha \int\frac{\rho_i(\vec{r}')-\rho_0(\vec{r}')}{|\vec{r}-\vec{r}'|}d\vec{r}' \notag \\
-\alpha^2 \Big\{\sum_{m=1}^{i-1} (1-\alpha)^{i-1-m}\int\frac{\rho_m(\vec{r}')-\rho_0(\vec{r}')}{|\vec{r}-\vec{r}'|}d\vec{r}' \label{zmp_eqn4}\Big\} \Big].
\end{eqnarray}
Therefore the contribution to $\Delta F$ from the term proportional to $\alpha$ (which is always positive) will be larger than that proportional to $\alpha^2$ (which could be positive or negative) if $\alpha <<1$ thereby ensuring $\Delta F \geq 0$ for each iterative step. This is seen to be the case while performing ZMP calculations.

\section{Summary}
\label{summary}
Exact results, whenever they can be obtained, are important to understand a theory properly.  This is particularly important in density functional theory since it is the most widely used theory of electronic structure but can be applied only approximately. For example, exact conditions on the exchange-correlation energy functionals have played an important role in their development. An important part of research in density functional theory has also been to construct the exact
Kohn-Sham system for known densities.  Not only the problem by itself is challenging, it also sets a benchmark against which approximate exchange-correlation functionals can be tested. 
\par Over the past thirty years or so, a variety of methods have been proposed to construct the Kohn-Sham system for a given density. These methods emerge from different ways of formulating the inverse problem. In this paper we have shown a majority of these methods (those formulated in terms of the density) to be results of the Euler equation for the density and have also provided an understanding of all these methods based on Lieb's definition of the Hohenberg-Kohn universal functional $F[\rho]$. Our work thus connects these different methods through a fundamental principle of DFT and gives a unified theory for the construction of Kohn-Sham system for a given density. As a result it also provides flexibility in ways through which the Kohn-Sham system can be constructed for a given density as has been demonstrated in the paper.

\section{Appendix: Derivation of the Kohn-Sham potential from Hartree-Fock wave function}
The LPS equation for HF density can be written as
\begin{equation}
[-\frac{1}{2}\nabla^2 + v_{eff}^{HF}(\vec{r})]\rho^{\frac{1}{2}}_{HF}(\vec{r})= \mu^{HF}\rho^{\frac{1}{2}}_{HF} \label{lps_hf}
\end{equation} 
where \cite{ Buijse_1989,Gritsenko_1994}

\begin{equation}
\begin{split}
v^{HF}_{eff}&=v_{ext}+v^{HF}_S+v^{HF}_H+  \frac{1}{\rho_{HF}} \sum_j (\epsilon_{max}^{HF}-\epsilon^{HF}_j)
|\phi_j^{HF}|^2  \\ 
&+\frac{1}{2} \sum_j \frac{|\nabla \phi^{HF}_j|^2}{\rho_{HF}} -\frac{1}{8} \frac{|\nabla \rho_{HF}|^2}{\rho^2_{HF}}.	\label{apn_veff_HF}
\end{split}
\end{equation} 
In the expression above $\phi^{HF}_j$ and $\epsilon^{HF}_j$ are the HF orbitals and their eigenenergies, respectively. The quantity $v^{HF}_S$ is the Slater potential \cite{Slater_1951} calculated from HF orbitals. \\
\par Similarly, for the Kohn-Sham equation, the effective potential \cite{LEVY_1988} for the corresponding LPS equation  is    
\begin{equation}
\begin{split}
v_{eff}^{KS}&=v_{ext}+v_H+ v_{x}+ \frac{1}{\rho_{KS}} \sum_j (\epsilon_{max}^{KS}-\epsilon^{KS}_j)
|\phi_j^{KS}|^2  \\ 
&+\frac{1}{2} \sum_j \frac{|\nabla \phi^{KS}_j|^2}{\rho_{KS}} -\frac{1}{8} \frac{|\nabla \rho_{KS}|^2}{\rho^2_{KS}}.	\label{apn_veff_ks}
\end{split}
\end{equation}
Now using $ \frac{1}{2}\frac{\nabla^2 \rho}{\rho}= \mu-v_{eff}$ to write Eq. \ref{master1} of the main text in terms of effective potentials, we get  
\begin{equation}
v_{xc}^{i+1}(\vec{r})=v_{xc}^i(\vec{r}) + v_{eff}^{WF}(\vec{r})- v^{i,KS}_{eff}(\vec{r}),\label{lps_itr1}
\end{equation}
where $v_{eff}^{WF}(\vec{r})$ is the effective potential for the interacting system. For the HF wavefunction its expression is given by Eq. \ref{apn_veff_HF}; the general expression for it is given in \cite{LPS_1984, Buijse_1989}. In addition, $v^{i,KS}_{eff}(\vec{r})$ is the  effective potential corresponding to non-interacting KS system at the $i^{th}$ iteration. In particular for HF wavefunction, the equation above becomes
\begin{equation}
\begin{split}
v^{i+1}_{x} &= \epsilon_{max}^{HF}-\epsilon_{max}^{i,KS}+ v^{HF}_S\\
&-\frac{\sum_j \epsilon^{HF}_j
	|\phi_j^{HF}|^2 }{\rho_{HF}} +\frac{\sum_j \epsilon^{i,KS}_j
	|\phi_j^{i,KS}|^2 }{\rho_{KS}^i} \\
&+ \frac{1}{2} \sum_j \frac{|\nabla \phi^{HF}_j|^2}{\rho_{HF}}-\frac{1}{2} \sum_j \frac{|\nabla \phi^{i,KS}_j|^2}{\rho_{KS}^i}. \label{viktor_exch}
\end{split}
\end{equation} 
In writing Eq. \ref{viktor_exch} all explicitly density dependent terms are canceled.  
\par Eq. \ref{viktor_exch} was first derived and implemented by  Ryabinkin, Kananenka and  Staroverov (RKS) \cite{Viktor_2013} to generate KS exchange potential corresponding to HF wavefunction generated from finite Gaussian basis set. This method gives approximate but highly accurate local exchange potential (essentially the same as the optimized effective potential) that is  free from unwanted oscillatory features that arise \cite{Schipper1997,Gidopoulos_2012,Viktor_2013} near the nucleus if Gaussian basis set is used to generate density. Staroverov and coworkers further extended it to many-body wavefunction \cite{Viktor_2015_JCP,Viktor_2015}. Eq. \ref{viktor_exch} was also used by Nagy \cite{Nagy_1997} to derive  the Krieger, Li, and Iafrate  approximation \cite{KLI_1992} to OEP.

\par An alternate method  to obtain the Kohn-Sham like non-interacting system for a given wave function has also been proposed in \cite{Gidopoulos_2011}. It has  recently been applied \cite{Hollins_2017} to obtain the local exchange potential for the Hartree-Fock wavefunction and the corresponding band-structure of solids. However the potential is generated using the Hartree-Fock density with $S[\rho]=\frac{\epsilon}{2}\iint\frac{\rho(\vec{r})\rho(\vec{r}')}{|\vec{r}-\vec{r}'|}d\vec{r}d\vec{r}' $  and a basis consisting of a large number of plane waves.

\bibliography{mybib}

\begin{thebibliography}{78}%
\makeatletter
\providecommand \@ifxundefined [1]{%
 \@ifx{#1\undefined}
}%
\providecommand \@ifnum [1]{%
 \ifnum #1\expandafter \@firstoftwo
 \else \expandafter \@secondoftwo
 \fi
}%
\providecommand \@ifx [1]{%
 \ifx #1\expandafter \@firstoftwo
 \else \expandafter \@secondoftwo
 \fi
}%
\providecommand \natexlab [1]{#1}%
\providecommand \enquote  [1]{``#1''}%
\providecommand \bibnamefont  [1]{#1}%
\providecommand \bibfnamefont [1]{#1}%
\providecommand \citenamefont [1]{#1}%
\providecommand \href@noop [0]{\@secondoftwo}%
\providecommand \href [0]{\begingroup \@sanitize@url \@href}%
\providecommand \@href[1]{\@@startlink{#1}\@@href}%
\providecommand \@@href[1]{\endgroup#1\@@endlink}%
\providecommand \@sanitize@url [0]{\catcode `\\12\catcode `\$12\catcode
  `\&12\catcode `\#12\catcode `\^12\catcode `\_12\catcode `\%12\relax}%
\providecommand \@@startlink[1]{}%
\providecommand \@@endlink[0]{}%
\providecommand \url  [0]{\begingroup\@sanitize@url \@url }%
\providecommand \@url [1]{\endgroup\@href {#1}{\urlprefix }}%
\providecommand \urlprefix  [0]{URL }%
\providecommand \Eprint [0]{\href }%
\providecommand \doibase [0]{http://dx.doi.org/}%
\providecommand \selectlanguage [0]{\@gobble}%
\providecommand \bibinfo  [0]{\@secondoftwo}%
\providecommand \bibfield  [0]{\@secondoftwo}%
\providecommand \translation [1]{[#1]}%
\providecommand \BibitemOpen [0]{}%
\providecommand \bibitemStop [0]{}%
\providecommand \bibitemNoStop [0]{.\EOS\space}%
\providecommand \EOS [0]{\spacefactor3000\relax}%
\providecommand \BibitemShut  [1]{\csname bibitem#1\endcsname}%
\let\auto@bib@innerbib\@empty
\bibitem [{\citenamefont {Parr}\ and\ \citenamefont {Yang}(1995)}]{Yang}%
  \BibitemOpen
  \bibfield  {author} {\bibinfo {author} {\bibfnamefont {R.~G.}\ \bibnamefont
  {Parr}}\ and\ \bibinfo {author} {\bibfnamefont {W.}~\bibnamefont {Yang}},\
  }\href@noop {} {\emph {\bibinfo {title} {Density-Functional Theory of Atoms
  and Molecules}}}\ (\bibinfo  {publisher} {Oxford Science Publications},\
  \bibinfo {year} {1995})\BibitemShut {NoStop}%
\bibitem [{\citenamefont {Engel}\ and\ \citenamefont {Dreizler}(2011)}]{Drei}%
  \BibitemOpen
  \bibfield  {author} {\bibinfo {author} {\bibfnamefont {E.}~\bibnamefont
  {Engel}}\ and\ \bibinfo {author} {\bibfnamefont {R.~M.}\ \bibnamefont
  {Dreizler}},\ }\href@noop {} {\emph {\bibinfo {title} {Density Functional
  Theory}}}\ (\bibinfo  {publisher} {Springer-Verlag Berlin Heidelberg},\
  \bibinfo {year} {2011})\BibitemShut {NoStop}%
\bibitem [{\citenamefont {Dreizler}\ and\ \citenamefont {Gross}(1990)}]{Gross}%
  \BibitemOpen
  \bibfield  {author} {\bibinfo {author} {\bibfnamefont {R.~M.}\ \bibnamefont
  {Dreizler}}\ and\ \bibinfo {author} {\bibfnamefont {E.~K.~U.}\ \bibnamefont
  {Gross}},\ }\href@noop {} {\emph {\bibinfo {title} {Density Functional
  Theory}}}\ (\bibinfo  {publisher} {Springer-Verlag Berlin Heidelberg},\
  \bibinfo {year} {1990})\BibitemShut {NoStop}%
\bibitem [{\citenamefont {Hohenberg}\ and\ \citenamefont
  {Kohn}(1964)}]{Hohenberg_PR.136.B864}%
  \BibitemOpen
  \bibfield  {author} {\bibinfo {author} {\bibfnamefont {P.}~\bibnamefont
  {Hohenberg}}\ and\ \bibinfo {author} {\bibfnamefont {W.}~\bibnamefont
  {Kohn}},\ }\href {https://link.aps.org/doi/10.1103/PhysRev.136.B864}
  {\bibfield  {journal} {\bibinfo  {journal} {Phys. Rev.}\ }\textbf {\bibinfo
  {volume} {136}},\ \bibinfo {pages} {B864} (\bibinfo {year}
  {1964})}\BibitemShut {NoStop}%
\bibitem [{\citenamefont {Jones}\ and\ \citenamefont
  {Gunnarsson}(1989)}]{JoneRMP_89}%
  \BibitemOpen
  \bibfield  {author} {\bibinfo {author} {\bibfnamefont {R.~O.}\ \bibnamefont
  {Jones}}\ and\ \bibinfo {author} {\bibfnamefont {O.}~\bibnamefont
  {Gunnarsson}},\ }\href {\doibase 10.1103/RevModPhys.61.689} {\bibfield
  {journal} {\bibinfo  {journal} {Rev. Mod. Phys.}\ }\textbf {\bibinfo {volume}
  {61}},\ \bibinfo {pages} {689} (\bibinfo {year} {1989})}\BibitemShut
  {NoStop}%
\bibitem [{\citenamefont {Spruch}(1991)}]{SpruchRMP_1991}%
  \BibitemOpen
  \bibfield  {author} {\bibinfo {author} {\bibfnamefont {L.}~\bibnamefont
  {Spruch}},\ }\href {\doibase 10.1103/RevModPhys.63.151} {\bibfield  {journal}
  {\bibinfo  {journal} {Rev. Mod. Phys.}\ }\textbf {\bibinfo {volume} {63}},\
  \bibinfo {pages} {151} (\bibinfo {year} {1991})}\BibitemShut {NoStop}%
\bibitem [{\citenamefont {Jones}(2015)}]{JoneRMP}%
  \BibitemOpen
  \bibfield  {author} {\bibinfo {author} {\bibfnamefont {R.~O.}\ \bibnamefont
  {Jones}},\ }\href {\doibase 10.1103/RevModPhys.87.897} {\bibfield  {journal}
  {\bibinfo  {journal} {Rev. Mod. Phys.}\ }\textbf {\bibinfo {volume} {87}},\
  \bibinfo {pages} {897} (\bibinfo {year} {2015})}\BibitemShut {NoStop}%
\bibitem [{\citenamefont {Sun}\ \emph {et~al.}(2016)\citenamefont {Sun},
  \citenamefont {Remsing}, \citenamefont {Zhang}, \citenamefont {Sun},
  \citenamefont {Ruzsinszky}, \citenamefont {Peng}, \citenamefont {Yang},
  \citenamefont {Paul}, \citenamefont {Waghmare}, \citenamefont {Wu},
  \citenamefont {Klein},\ and\ \citenamefont {Perdew}}]{SCANACC}%
  \BibitemOpen
  \bibfield  {author} {\bibinfo {author} {\bibfnamefont {J.}~\bibnamefont
  {Sun}}, \bibinfo {author} {\bibfnamefont {R.~C.}\ \bibnamefont {Remsing}},
  \bibinfo {author} {\bibfnamefont {Y.}~\bibnamefont {Zhang}}, \bibinfo
  {author} {\bibfnamefont {Z.}~\bibnamefont {Sun}}, \bibinfo {author}
  {\bibfnamefont {A.}~\bibnamefont {Ruzsinszky}}, \bibinfo {author}
  {\bibfnamefont {H.}~\bibnamefont {Peng}}, \bibinfo {author} {\bibfnamefont
  {Z.}~\bibnamefont {Yang}}, \bibinfo {author} {\bibfnamefont {A.}~\bibnamefont
  {Paul}}, \bibinfo {author} {\bibfnamefont {U.}~\bibnamefont {Waghmare}},
  \bibinfo {author} {\bibfnamefont {X.}~\bibnamefont {Wu}}, \bibinfo {author}
  {\bibfnamefont {M.~L.}\ \bibnamefont {Klein}}, \ and\ \bibinfo {author}
  {\bibfnamefont {J.~P.}\ \bibnamefont {Perdew}},\ }\href
  {http://dx.doi.org/10.1038/nchem.2535} {\bibfield  {journal} {\bibinfo
  {journal} {Nat. Chem.}\ }\textbf {\bibinfo {volume} {8}},\ \bibinfo {pages}
  {831} (\bibinfo {year} {2016})}\BibitemShut {NoStop}%
\bibitem [{\citenamefont {Kohn}\ and\ \citenamefont
  {Sham}(1965)}]{Kohn_PR.140.A1133}%
  \BibitemOpen
  \bibfield  {author} {\bibinfo {author} {\bibfnamefont {W.}~\bibnamefont
  {Kohn}}\ and\ \bibinfo {author} {\bibfnamefont {L.~J.}\ \bibnamefont
  {Sham}},\ }\href {https://link.aps.org/doi/10.1103/PhysRev.140.A1133}
  {\bibfield  {journal} {\bibinfo  {journal} {Phys. Rev.}\ }\textbf {\bibinfo
  {volume} {140}},\ \bibinfo {pages} {A1133} (\bibinfo {year}
  {1965})}\BibitemShut {NoStop}%
\bibitem [{\citenamefont {Perdew}\ \emph {et~al.}(1982)\citenamefont {Perdew},
  \citenamefont {Parr}, \citenamefont {Levy},\ and\ \citenamefont {Jr}}]{PPLB}%
  \BibitemOpen
  \bibfield  {author} {\bibinfo {author} {\bibfnamefont {J.~P.}\ \bibnamefont
  {Perdew}}, \bibinfo {author} {\bibfnamefont {R.~G.}\ \bibnamefont {Parr}},
  \bibinfo {author} {\bibfnamefont {M.}~\bibnamefont {Levy}}, \ and\ \bibinfo
  {author} {\bibfnamefont {J.~L.~B.}\ \bibnamefont {Jr}},\ }\href
  {http://journals.aps.org/prl/pdf/10.1103/PhysRevLett.49.1691} {\bibfield
  {journal} {\bibinfo  {journal} {Phys. Rev. Lett.}\ }\textbf {\bibinfo
  {volume} {49}},\ \bibinfo {pages} {1691} (\bibinfo {year}
  {1982})}\BibitemShut {NoStop}%
\bibitem [{\citenamefont {Medvedev}\ \emph {et~al.}(2017)\citenamefont
  {Medvedev}, \citenamefont {Bushmarinov}, \citenamefont {Sun}, \citenamefont
  {Perdew},\ and\ \citenamefont {Lyssenko}}]{Medv_2017}%
  \BibitemOpen
  \bibfield  {author} {\bibinfo {author} {\bibfnamefont {M.~G.}\ \bibnamefont
  {Medvedev}}, \bibinfo {author} {\bibfnamefont {I.~S.}\ \bibnamefont
  {Bushmarinov}}, \bibinfo {author} {\bibfnamefont {J.}~\bibnamefont {Sun}},
  \bibinfo {author} {\bibfnamefont {J.~P.}\ \bibnamefont {Perdew}}, \ and\
  \bibinfo {author} {\bibfnamefont {K.~A.}\ \bibnamefont {Lyssenko}},\ }\href
  {\doibase 10.1126/science.aah5975} {\bibfield  {journal} {\bibinfo  {journal}
  {Science}\ }\textbf {\bibinfo {volume} {355}},\ \bibinfo {pages} {49}
  (\bibinfo {year} {2017})}\BibitemShut {NoStop}%
\bibitem [{\citenamefont {Peverati}\ and\ \citenamefont
  {Truhlar}(2014)}]{TruhF}%
  \BibitemOpen
  \bibfield  {author} {\bibinfo {author} {\bibfnamefont {R.}~\bibnamefont
  {Peverati}}\ and\ \bibinfo {author} {\bibfnamefont {D.~G.}\ \bibnamefont
  {Truhlar}},\ }\href {\doibase 10.1098/rsta.2012.0476} {\bibfield  {journal}
  {\bibinfo  {journal} {Philosophical Transactions of the Royal Society of
  London A: Mathematical, Physical and Engineering Sciences}\ }\textbf
  {\bibinfo {volume} {372}} (\bibinfo {year} {2014}),\
  10.1098/rsta.2012.0476},\ \Eprint
  {http://arxiv.org/abs/http://rsta.royalsocietypublishing.org/content/372/2011/20120476.full.pdf}
  {http://rsta.royalsocietypublishing.org/content/372/2011/20120476.full.pdf}
  \BibitemShut {NoStop}%
\bibitem [{\citenamefont {Perdew}\ \emph {et~al.}(1996)\citenamefont {Perdew},
  \citenamefont {Burke},\ and\ \citenamefont {Ernzerhof}}]{Perdew_PRL.77.3865}%
  \BibitemOpen
  \bibfield  {author} {\bibinfo {author} {\bibfnamefont {J.~P.}\ \bibnamefont
  {Perdew}}, \bibinfo {author} {\bibfnamefont {K.}~\bibnamefont {Burke}}, \
  and\ \bibinfo {author} {\bibfnamefont {M.}~\bibnamefont {Ernzerhof}},\ }\href
  {https://link.aps.org/doi/10.1103/PhysRevLett.77.3865} {\bibfield  {journal}
  {\bibinfo  {journal} {Phys. Rev. Lett.}\ }\textbf {\bibinfo {volume} {77}},\
  \bibinfo {pages} {3865} (\bibinfo {year} {1996})}\BibitemShut {NoStop}%
\bibitem [{\citenamefont {Perdew}\ \emph {et~al.}(1999)\citenamefont {Perdew},
  \citenamefont {Kurth}, \citenamefont {Zupan},\ and\ \citenamefont
  {Blaha}}]{Perdew_PRL.82.2544}%
  \BibitemOpen
  \bibfield  {author} {\bibinfo {author} {\bibfnamefont {J.~P.}\ \bibnamefont
  {Perdew}}, \bibinfo {author} {\bibfnamefont {S.}~\bibnamefont {Kurth}},
  \bibinfo {author} {\bibfnamefont {A.}~\bibnamefont {Zupan}}, \ and\ \bibinfo
  {author} {\bibfnamefont {P.}~\bibnamefont {Blaha}},\ }\href
  {https://link.aps.org/doi/10.1103/PhysRevLett.82.2544} {\bibfield  {journal}
  {\bibinfo  {journal} {Phys. Rev. Lett.}\ }\textbf {\bibinfo {volume} {82}},\
  \bibinfo {pages} {2544} (\bibinfo {year} {1999})}\BibitemShut {NoStop}%
\bibitem [{\citenamefont {Tao}\ \emph {et~al.}(2003)\citenamefont {Tao},
  \citenamefont {Perdew}, \citenamefont {Staroverov},\ and\ \citenamefont
  {Scuseria}}]{Jianmin_PRL.91.146401}%
  \BibitemOpen
  \bibfield  {author} {\bibinfo {author} {\bibfnamefont {J.}~\bibnamefont
  {Tao}}, \bibinfo {author} {\bibfnamefont {J.~P.}\ \bibnamefont {Perdew}},
  \bibinfo {author} {\bibfnamefont {V.~N.}\ \bibnamefont {Staroverov}}, \ and\
  \bibinfo {author} {\bibfnamefont {G.~E.}\ \bibnamefont {Scuseria}},\ }\href
  {https://link.aps.org/doi/10.1103/PhysRevLett.91.146401} {\bibfield
  {journal} {\bibinfo  {journal} {Phys. Rev. Lett.}\ }\textbf {\bibinfo
  {volume} {91}},\ \bibinfo {pages} {146401} (\bibinfo {year}
  {2003})}\BibitemShut {NoStop}%
\bibitem [{\citenamefont {Sun}\ \emph {et~al.}(2015{\natexlab{a}})\citenamefont
  {Sun}, \citenamefont {Perdew},\ and\ \citenamefont {Ruzsinszky}}]{Sun_PNAS}%
  \BibitemOpen
  \bibfield  {author} {\bibinfo {author} {\bibfnamefont {J.}~\bibnamefont
  {Sun}}, \bibinfo {author} {\bibfnamefont {J.~P.}\ \bibnamefont {Perdew}}, \
  and\ \bibinfo {author} {\bibfnamefont {A.}~\bibnamefont {Ruzsinszky}},\
  }\href {\doibase 10.1073/pnas.1423145112} {\bibfield  {journal} {\bibinfo
  {journal} {Proc. Natl. Acad. Sci. U.S.A}\ }\textbf {\bibinfo {volume}
  {112}},\ \bibinfo {pages} {685} (\bibinfo {year}
  {2015}{\natexlab{a}})}\BibitemShut {NoStop}%
\bibitem [{\citenamefont {Sun}\ \emph {et~al.}(2015{\natexlab{b}})\citenamefont
  {Sun}, \citenamefont {Ruzsinszky},\ and\ \citenamefont
  {Perdew}}]{Sun_PRL.115.036402}%
  \BibitemOpen
  \bibfield  {author} {\bibinfo {author} {\bibfnamefont {J.}~\bibnamefont
  {Sun}}, \bibinfo {author} {\bibfnamefont {A.}~\bibnamefont {Ruzsinszky}}, \
  and\ \bibinfo {author} {\bibfnamefont {J.~P.}\ \bibnamefont {Perdew}},\
  }\href {https://link.aps.org/doi/10.1103/PhysRevLett.115.036402} {\bibfield
  {journal} {\bibinfo  {journal} {Phys. Rev. Lett.}\ }\textbf {\bibinfo
  {volume} {115}},\ \bibinfo {pages} {036402} (\bibinfo {year}
  {2015}{\natexlab{b}})}\BibitemShut {NoStop}%
\bibitem [{\citenamefont {Becke}(1993)}]{Becke_JCP.98.5648}%
  \BibitemOpen
  \bibfield  {author} {\bibinfo {author} {\bibfnamefont {A.~D.}\ \bibnamefont
  {Becke}},\ }\href {http://dx.doi.org/10.1063/1.464913} {\bibfield  {journal}
  {\bibinfo  {journal} {J. Chem. Phys.}\ }\textbf {\bibinfo {volume} {98}},\
  \bibinfo {pages} {5648} (\bibinfo {year} {1993})}\BibitemShut {NoStop}%
\bibitem [{\citenamefont {Lee}\ \emph {et~al.}(1988)\citenamefont {Lee},
  \citenamefont {Yang},\ and\ \citenamefont {Parr}}]{Lee_PRB.37.785}%
  \BibitemOpen
  \bibfield  {author} {\bibinfo {author} {\bibfnamefont {C.}~\bibnamefont
  {Lee}}, \bibinfo {author} {\bibfnamefont {W.}~\bibnamefont {Yang}}, \ and\
  \bibinfo {author} {\bibfnamefont {R.~G.}\ \bibnamefont {Parr}},\ }\href
  {https://link.aps.org/doi/10.1103/PhysRevB.37.785} {\bibfield  {journal}
  {\bibinfo  {journal} {Phys. Rev. B}\ }\textbf {\bibinfo {volume} {37}},\
  \bibinfo {pages} {785} (\bibinfo {year} {1988})}\BibitemShut {NoStop}%
\bibitem [{\citenamefont {Vosko}\ \emph {et~al.}(1980)\citenamefont {Vosko},
  \citenamefont {Wilk},\ and\ \citenamefont {Nusair}}]{Vosko_CJP.58.1200}%
  \BibitemOpen
  \bibfield  {author} {\bibinfo {author} {\bibfnamefont {S.~H.}\ \bibnamefont
  {Vosko}}, \bibinfo {author} {\bibfnamefont {L.}~\bibnamefont {Wilk}}, \ and\
  \bibinfo {author} {\bibfnamefont {M.}~\bibnamefont {Nusair}},\ }\href
  {https://doi.org/10.1139/p80-159} {\bibfield  {journal} {\bibinfo  {journal}
  {Canadian Journal of Physics}\ }\textbf {\bibinfo {volume} {58}},\ \bibinfo
  {pages} {1200} (\bibinfo {year} {1980})}\BibitemShut {NoStop}%
\bibitem [{\citenamefont {Stephens}\ \emph {et~al.}(1994)\citenamefont
  {Stephens}, \citenamefont {Devlin}, \citenamefont {Chabalowski},\ and\
  \citenamefont {Frisch}}]{Stephens_JPC.98.11623}%
  \BibitemOpen
  \bibfield  {author} {\bibinfo {author} {\bibfnamefont {P.~J.}\ \bibnamefont
  {Stephens}}, \bibinfo {author} {\bibfnamefont {F.~J.}\ \bibnamefont
  {Devlin}}, \bibinfo {author} {\bibfnamefont {C.~F.}\ \bibnamefont
  {Chabalowski}}, \ and\ \bibinfo {author} {\bibfnamefont {M.~J.}\ \bibnamefont
  {Frisch}},\ }\href {http://dx.doi.org/10.1021/j100096a001} {\bibfield
  {journal} {\bibinfo  {journal} {J. Phys. Chem.}\ }\textbf {\bibinfo {volume}
  {98}},\ \bibinfo {pages} {11623} (\bibinfo {year} {1994})}\BibitemShut
  {NoStop}%
\bibitem [{\citenamefont {Chen}\ \emph {et~al.}(2017)\citenamefont {Chen},
  \citenamefont {Ko}, \citenamefont {Remsing}, \citenamefont
  {Calegari~Andrade}, \citenamefont {Santra}, \citenamefont {Sun},
  \citenamefont {Selloni}, \citenamefont {Car}, \citenamefont {Klein},
  \citenamefont {Perdew},\ and\ \citenamefont {Wu}}]{Chen_2017}%
  \BibitemOpen
  \bibfield  {author} {\bibinfo {author} {\bibfnamefont {M.}~\bibnamefont
  {Chen}}, \bibinfo {author} {\bibfnamefont {H.-Y.}\ \bibnamefont {Ko}},
  \bibinfo {author} {\bibfnamefont {R.~C.}\ \bibnamefont {Remsing}}, \bibinfo
  {author} {\bibfnamefont {M.~F.}\ \bibnamefont {Calegari~Andrade}}, \bibinfo
  {author} {\bibfnamefont {B.}~\bibnamefont {Santra}}, \bibinfo {author}
  {\bibfnamefont {Z.}~\bibnamefont {Sun}}, \bibinfo {author} {\bibfnamefont
  {A.}~\bibnamefont {Selloni}}, \bibinfo {author} {\bibfnamefont
  {R.}~\bibnamefont {Car}}, \bibinfo {author} {\bibfnamefont {M.~L.}\
  \bibnamefont {Klein}}, \bibinfo {author} {\bibfnamefont {J.~P.}\ \bibnamefont
  {Perdew}}, \ and\ \bibinfo {author} {\bibfnamefont {X.}~\bibnamefont {Wu}},\
  }\href {\doibase 10.1073/pnas.1712499114} {\bibfield  {journal} {\bibinfo
  {journal} {Proc. Natl. Acad. Sci. U.S.A}\ }\textbf {\bibinfo {volume}
  {114}},\ \bibinfo {pages} {10846} (\bibinfo {year} {2017})}\BibitemShut
  {NoStop}%
\bibitem [{\citenamefont {Aryasetiawan}\ and\ \citenamefont
  {Stott}(1988)}]{Stott_1988}%
  \BibitemOpen
  \bibfield  {author} {\bibinfo {author} {\bibfnamefont {F.}~\bibnamefont
  {Aryasetiawan}}\ and\ \bibinfo {author} {\bibfnamefont {M.~J.}\ \bibnamefont
  {Stott}},\ }\href {\doibase 10.1103/PhysRevB.38.2974} {\bibfield  {journal}
  {\bibinfo  {journal} {Phys. Rev. B}\ }\textbf {\bibinfo {volume} {38}},\
  \bibinfo {pages} {2974} (\bibinfo {year} {1988})}\BibitemShut {NoStop}%
\bibitem [{\citenamefont {G\"orling}(1992)}]{Gorling_1992}%
  \BibitemOpen
  \bibfield  {author} {\bibinfo {author} {\bibfnamefont {A.}~\bibnamefont
  {G\"orling}},\ }\href {\doibase 10.1103/PhysRevA.46.3753} {\bibfield
  {journal} {\bibinfo  {journal} {Phys. Rev. A}\ }\textbf {\bibinfo {volume}
  {46}},\ \bibinfo {pages} {3753} (\bibinfo {year} {1992})}\BibitemShut
  {NoStop}%
\bibitem [{\citenamefont {Zhao}\ and\ \citenamefont {Parr}(1992)}]{Zhao_1992}%
  \BibitemOpen
  \bibfield  {author} {\bibinfo {author} {\bibfnamefont {Q.}~\bibnamefont
  {Zhao}}\ and\ \bibinfo {author} {\bibfnamefont {R.~G.}\ \bibnamefont
  {Parr}},\ }\href {\doibase 10.1103/PhysRevA.46.2337} {\bibfield  {journal}
  {\bibinfo  {journal} {Phys. Rev. A}\ }\textbf {\bibinfo {volume} {46}},\
  \bibinfo {pages} {2337} (\bibinfo {year} {1992})}\BibitemShut {NoStop}%
\bibitem [{\citenamefont {Zhao}\ and\ \citenamefont {Parr}(1993)}]{Zhao_1993}%
  \BibitemOpen
  \bibfield  {author} {\bibinfo {author} {\bibfnamefont {Q.}~\bibnamefont
  {Zhao}}\ and\ \bibinfo {author} {\bibfnamefont {R.~G.}\ \bibnamefont
  {Parr}},\ }\href {\doibase 10.1063/1.465093} {\bibfield  {journal} {\bibinfo
  {journal} {J. Chem. Phys.}\ }\textbf {\bibinfo {volume} {98}},\ \bibinfo
  {pages} {543} (\bibinfo {year} {1993})}\BibitemShut {NoStop}%
\bibitem [{\citenamefont {Wang}\ and\ \citenamefont {Parr}(1993)}]{Wang_1993}%
  \BibitemOpen
  \bibfield  {author} {\bibinfo {author} {\bibfnamefont {Y.}~\bibnamefont
  {Wang}}\ and\ \bibinfo {author} {\bibfnamefont {R.~G.}\ \bibnamefont
  {Parr}},\ }\href {\doibase 10.1103/PhysRevA.47.R1591} {\bibfield  {journal}
  {\bibinfo  {journal} {Phys. Rev. A}\ }\textbf {\bibinfo {volume} {47}},\
  \bibinfo {pages} {R1591} (\bibinfo {year} {1993})}\BibitemShut {NoStop}%
\bibitem [{\citenamefont {Zhao}\ \emph {et~al.}(1994)\citenamefont {Zhao},
  \citenamefont {Morrison},\ and\ \citenamefont {Parr}}]{Zhao_1994}%
  \BibitemOpen
  \bibfield  {author} {\bibinfo {author} {\bibfnamefont {Q.}~\bibnamefont
  {Zhao}}, \bibinfo {author} {\bibfnamefont {R.~C.}\ \bibnamefont {Morrison}},
  \ and\ \bibinfo {author} {\bibfnamefont {R.~G.}\ \bibnamefont {Parr}},\
  }\href {\doibase 10.1103/PhysRevA.50.2138} {\bibfield  {journal} {\bibinfo
  {journal} {Phys. Rev. A}\ }\textbf {\bibinfo {volume} {50}},\ \bibinfo
  {pages} {2138} (\bibinfo {year} {1994})}\BibitemShut {NoStop}%
\bibitem [{\citenamefont {van Leeuwen}\ and\ \citenamefont
  {Baerends}(1994)}]{Vlb_1994}%
  \BibitemOpen
  \bibfield  {author} {\bibinfo {author} {\bibfnamefont {R.}~\bibnamefont {van
  Leeuwen}}\ and\ \bibinfo {author} {\bibfnamefont {E.~J.}\ \bibnamefont
  {Baerends}},\ }\href {\doibase 10.1103/PhysRevA.49.2421} {\bibfield
  {journal} {\bibinfo  {journal} {Phys. Rev. A}\ }\textbf {\bibinfo {volume}
  {49}},\ \bibinfo {pages} {2421} (\bibinfo {year} {1994})}\BibitemShut
  {NoStop}%
\bibitem [{\citenamefont {Yang}\ and\ \citenamefont {Wu}(2002)}]{WY1}%
  \BibitemOpen
  \bibfield  {author} {\bibinfo {author} {\bibfnamefont {W.}~\bibnamefont
  {Yang}}\ and\ \bibinfo {author} {\bibfnamefont {Q.}~\bibnamefont {Wu}},\
  }\href {\doibase 10.1103/PhysRevLett.89.143002} {\bibfield  {journal}
  {\bibinfo  {journal} {Phys. Rev. Lett.}\ }\textbf {\bibinfo {volume} {89}},\
  \bibinfo {pages} {143002} (\bibinfo {year} {2002})}\BibitemShut {NoStop}%
\bibitem [{\citenamefont {Wu}\ and\ \citenamefont {Yang}(2003)}]{WY2}%
  \BibitemOpen
  \bibfield  {author} {\bibinfo {author} {\bibfnamefont {Q.}~\bibnamefont
  {Wu}}\ and\ \bibinfo {author} {\bibfnamefont {W.}~\bibnamefont {Yang}},\
  }\href {\doibase 10.1063/1.1535422} {\bibfield  {journal} {\bibinfo
  {journal} {J. Chem. Phys.}\ }\textbf {\bibinfo {volume} {118}},\ \bibinfo
  {pages} {2498} (\bibinfo {year} {2003})}\BibitemShut {NoStop}%
\bibitem [{\citenamefont {Peirs}\ \emph {et~al.}(2003)\citenamefont {Peirs},
  \citenamefont {Van~Neck},\ and\ \citenamefont {Waroquier}}]{Peirs_2003}%
  \BibitemOpen
  \bibfield  {author} {\bibinfo {author} {\bibfnamefont {K.}~\bibnamefont
  {Peirs}}, \bibinfo {author} {\bibfnamefont {D.}~\bibnamefont {Van~Neck}}, \
  and\ \bibinfo {author} {\bibfnamefont {M.}~\bibnamefont {Waroquier}},\ }\href
  {\doibase 10.1103/PhysRevA.67.012505} {\bibfield  {journal} {\bibinfo
  {journal} {Phys. Rev. A}\ }\textbf {\bibinfo {volume} {67}},\ \bibinfo
  {pages} {012505} (\bibinfo {year} {2003})}\BibitemShut {NoStop}%
\bibitem [{\citenamefont {Kadantsev}\ and\ \citenamefont
  {Stott}(2004)}]{Stott_2004}%
  \BibitemOpen
  \bibfield  {author} {\bibinfo {author} {\bibfnamefont {E.~S.}\ \bibnamefont
  {Kadantsev}}\ and\ \bibinfo {author} {\bibfnamefont {M.~J.}\ \bibnamefont
  {Stott}},\ }\href {\doibase 10.1103/PhysRevA.69.012502} {\bibfield  {journal}
  {\bibinfo  {journal} {Phys. Rev. A}\ }\textbf {\bibinfo {volume} {69}},\
  \bibinfo {pages} {012502} (\bibinfo {year} {2004})}\BibitemShut {NoStop}%
\bibitem [{\citenamefont {Ryabinkin}\ and\ \citenamefont
  {Staroverov}(2012)}]{Viktor_2012}%
  \BibitemOpen
  \bibfield  {author} {\bibinfo {author} {\bibfnamefont {I.~G.}\ \bibnamefont
  {Ryabinkin}}\ and\ \bibinfo {author} {\bibfnamefont {V.~N.}\ \bibnamefont
  {Staroverov}},\ }\href {\doibase 10.1063/1.4763481} {\bibfield  {journal}
  {\bibinfo  {journal} {J. Chem. Phys.}\ }\textbf {\bibinfo {volume} {137}},\
  \bibinfo {pages} {164113} (\bibinfo {year} {2012})}\BibitemShut {NoStop}%
\bibitem [{\citenamefont {Ryabinkin}\ \emph {et~al.}(2013)\citenamefont
  {Ryabinkin}, \citenamefont {Kananenka},\ and\ \citenamefont
  {Staroverov}}]{Viktor_2013}%
  \BibitemOpen
  \bibfield  {author} {\bibinfo {author} {\bibfnamefont {I.~G.}\ \bibnamefont
  {Ryabinkin}}, \bibinfo {author} {\bibfnamefont {A.~A.}\ \bibnamefont
  {Kananenka}}, \ and\ \bibinfo {author} {\bibfnamefont {V.~N.}\ \bibnamefont
  {Staroverov}},\ }\href {\doibase 10.1103/PhysRevLett.111.013001} {\bibfield
  {journal} {\bibinfo  {journal} {Phys. Rev. Lett.}\ }\textbf {\bibinfo
  {volume} {111}},\ \bibinfo {pages} {013001} (\bibinfo {year}
  {2013})}\BibitemShut {NoStop}%
\bibitem [{\citenamefont {Wagner}\ \emph {et~al.}(2014)\citenamefont {Wagner},
  \citenamefont {Baker}, \citenamefont {Stoudenmire}, \citenamefont {Burke},\
  and\ \citenamefont {White}}]{Wagner_2014}%
  \BibitemOpen
  \bibfield  {author} {\bibinfo {author} {\bibfnamefont {L.~O.}\ \bibnamefont
  {Wagner}}, \bibinfo {author} {\bibfnamefont {T.~E.}\ \bibnamefont {Baker}},
  \bibinfo {author} {\bibfnamefont {E.~M.}\ \bibnamefont {Stoudenmire}},
  \bibinfo {author} {\bibfnamefont {K.}~\bibnamefont {Burke}}, \ and\ \bibinfo
  {author} {\bibfnamefont {S.~R.}\ \bibnamefont {White}},\ }\href {\doibase
  10.1103/PhysRevB.90.045109} {\bibfield  {journal} {\bibinfo  {journal} {Phys.
  Rev. B}\ }\textbf {\bibinfo {volume} {90}},\ \bibinfo {pages} {045109}
  (\bibinfo {year} {2014})}\BibitemShut {NoStop}%
\bibitem [{\citenamefont {Ryabinkin}\ \emph {et~al.}(2015)\citenamefont
  {Ryabinkin}, \citenamefont {Kohut},\ and\ \citenamefont
  {Staroverov}}]{Viktor_2015}%
  \BibitemOpen
  \bibfield  {author} {\bibinfo {author} {\bibfnamefont {I.~G.}\ \bibnamefont
  {Ryabinkin}}, \bibinfo {author} {\bibfnamefont {S.~V.}\ \bibnamefont
  {Kohut}}, \ and\ \bibinfo {author} {\bibfnamefont {V.~N.}\ \bibnamefont
  {Staroverov}},\ }\href {\doibase 10.1103/PhysRevLett.115.083001} {\bibfield
  {journal} {\bibinfo  {journal} {Phys. Rev. Lett.}\ }\textbf {\bibinfo
  {volume} {115}},\ \bibinfo {pages} {083001} (\bibinfo {year}
  {2015})}\BibitemShut {NoStop}%
\bibitem [{\citenamefont {Jensen}\ and\ \citenamefont
  {Wasserman}(2017)}]{Wasserman_2017}%
  \BibitemOpen
  \bibfield  {author} {\bibinfo {author} {\bibfnamefont {D.~S.}\ \bibnamefont
  {Jensen}}\ and\ \bibinfo {author} {\bibfnamefont {A.}~\bibnamefont
  {Wasserman}},\ }\href {http://dx.doi.org/10.1002/qua.25425} {\bibfield
  {journal} {\bibinfo  {journal} {Int. J. Quantum Chem.}\ } (\bibinfo {year}
  {2017})}\BibitemShut {NoStop}%
\bibitem [{\citenamefont {Buijse}\ \emph {et~al.}(1989)\citenamefont {Buijse},
  \citenamefont {Baerends},\ and\ \citenamefont {Snijders}}]{Buijse_1989}%
  \BibitemOpen
  \bibfield  {author} {\bibinfo {author} {\bibfnamefont {M.~A.}\ \bibnamefont
  {Buijse}}, \bibinfo {author} {\bibfnamefont {E.~J.}\ \bibnamefont
  {Baerends}}, \ and\ \bibinfo {author} {\bibfnamefont {J.~G.}\ \bibnamefont
  {Snijders}},\ }\href {\doibase 10.1103/PhysRevA.40.4190} {\bibfield
  {journal} {\bibinfo  {journal} {Phys. Rev. A}\ }\textbf {\bibinfo {volume}
  {40}},\ \bibinfo {pages} {4190} (\bibinfo {year} {1989})}\BibitemShut
  {NoStop}%
\bibitem [{\citenamefont {Gritsenko}\ and\ \citenamefont
  {Baerends}(1996)}]{Gritsenko_1996}%
  \BibitemOpen
  \bibfield  {author} {\bibinfo {author} {\bibfnamefont {O.~V.}\ \bibnamefont
  {Gritsenko}}\ and\ \bibinfo {author} {\bibfnamefont {E.~J.}\ \bibnamefont
  {Baerends}},\ }\href {\doibase 10.1103/PhysRevA.54.1957} {\bibfield
  {journal} {\bibinfo  {journal} {Phys. Rev. A}\ }\textbf {\bibinfo {volume}
  {54}},\ \bibinfo {pages} {1957} (\bibinfo {year} {1996})}\BibitemShut
  {NoStop}%
\bibitem [{\citenamefont {Teale}\ \emph {et~al.}(2009)\citenamefont {Teale},
  \citenamefont {Coriani},\ and\ \citenamefont {Helgaker}}]{Teal2}%
  \BibitemOpen
  \bibfield  {author} {\bibinfo {author} {\bibfnamefont {A.~M.}\ \bibnamefont
  {Teale}}, \bibinfo {author} {\bibfnamefont {S.}~\bibnamefont {Coriani}}, \
  and\ \bibinfo {author} {\bibfnamefont {T.}~\bibnamefont {Helgaker}},\ }\href
  {\doibase 10.1063/1.3082285} {\bibfield  {journal} {\bibinfo  {journal} {J.
  Chem. Phys.}\ }\textbf {\bibinfo {volume} {130}},\ \bibinfo {pages} {104111}
  (\bibinfo {year} {2009})}\BibitemShut {NoStop}%
\bibitem [{\citenamefont {Teale}\ \emph
  {et~al.}(2010{\natexlab{a}})\citenamefont {Teale}, \citenamefont {Coriani},\
  and\ \citenamefont {Helgaker}}]{Teal3}%
  \BibitemOpen
  \bibfield  {author} {\bibinfo {author} {\bibfnamefont {A.~M.}\ \bibnamefont
  {Teale}}, \bibinfo {author} {\bibfnamefont {S.}~\bibnamefont {Coriani}}, \
  and\ \bibinfo {author} {\bibfnamefont {T.}~\bibnamefont {Helgaker}},\ }\href
  {\doibase 10.1063/1.3380834} {\bibfield  {journal} {\bibinfo  {journal} {J.
  Chem. Phys.}\ }\textbf {\bibinfo {volume} {132}},\ \bibinfo {pages} {164115}
  (\bibinfo {year} {2010}{\natexlab{a}})}\BibitemShut {NoStop}%
\bibitem [{\citenamefont {Teale}\ \emph
  {et~al.}(2010{\natexlab{b}})\citenamefont {Teale}, \citenamefont {Coriani},\
  and\ \citenamefont {Helgaker}}]{Teal4}%
  \BibitemOpen
  \bibfield  {author} {\bibinfo {author} {\bibfnamefont {A.~M.}\ \bibnamefont
  {Teale}}, \bibinfo {author} {\bibfnamefont {S.}~\bibnamefont {Coriani}}, \
  and\ \bibinfo {author} {\bibfnamefont {T.}~\bibnamefont {Helgaker}},\ }\href
  {\doibase 10.1063/1.3488100} {\bibfield  {journal} {\bibinfo  {journal} {J.
  Chem. Phys.}\ }\textbf {\bibinfo {volume} {133}},\ \bibinfo {pages} {164112}
  (\bibinfo {year} {2010}{\natexlab{b}})}\BibitemShut {NoStop}%
\bibitem [{\citenamefont {Makmal}\ \emph {et~al.}(2011)\citenamefont {Makmal},
  \citenamefont {K\"ummel},\ and\ \citenamefont {Kronik}}]{Makmal_2011}%
  \BibitemOpen
  \bibfield  {author} {\bibinfo {author} {\bibfnamefont {A.}~\bibnamefont
  {Makmal}}, \bibinfo {author} {\bibfnamefont {S.}~\bibnamefont {K\"ummel}}, \
  and\ \bibinfo {author} {\bibfnamefont {L.}~\bibnamefont {Kronik}},\ }\href
  {\doibase 10.1103/PhysRevA.83.062512} {\bibfield  {journal} {\bibinfo
  {journal} {Phys. Rev. A}\ }\textbf {\bibinfo {volume} {83}},\ \bibinfo
  {pages} {062512} (\bibinfo {year} {2011})}\BibitemShut {NoStop}%
\bibitem [{\citenamefont {Stoudenmire}\ \emph {et~al.}(2012)\citenamefont
  {Stoudenmire}, \citenamefont {Wagner}, \citenamefont {White},\ and\
  \citenamefont {Burke}}]{Wagner_2012}%
  \BibitemOpen
  \bibfield  {author} {\bibinfo {author} {\bibfnamefont {E.~M.}\ \bibnamefont
  {Stoudenmire}}, \bibinfo {author} {\bibfnamefont {L.~O.}\ \bibnamefont
  {Wagner}}, \bibinfo {author} {\bibfnamefont {S.~R.}\ \bibnamefont {White}}, \
  and\ \bibinfo {author} {\bibfnamefont {K.}~\bibnamefont {Burke}},\ }\href
  {\doibase 10.1103/PhysRevLett.109.056402} {\bibfield  {journal} {\bibinfo
  {journal} {Phys. Rev. Lett.}\ }\textbf {\bibinfo {volume} {109}},\ \bibinfo
  {pages} {056402} (\bibinfo {year} {2012})}\BibitemShut {NoStop}%
\bibitem [{\citenamefont {Kohut}\ \emph {et~al.}(2016)\citenamefont {Kohut},
  \citenamefont {Polgar},\ and\ \citenamefont {Staroverov}}]{Kohut_2016}%
  \BibitemOpen
  \bibfield  {author} {\bibinfo {author} {\bibfnamefont {S.~V.}\ \bibnamefont
  {Kohut}}, \bibinfo {author} {\bibfnamefont {A.~M.}\ \bibnamefont {Polgar}}, \
  and\ \bibinfo {author} {\bibfnamefont {V.~N.}\ \bibnamefont {Staroverov}},\
  }\href {\doibase 10.1039/C6CP00878J} {\bibfield  {journal} {\bibinfo
  {journal} {Phys. Chem. Chem. Phys.}\ }\textbf {\bibinfo {volume} {18}},\
  \bibinfo {pages} {20938} (\bibinfo {year} {2016})}\BibitemShut {NoStop}%
\bibitem [{\citenamefont {Ben\'{\i}tez}\ and\ \citenamefont
  {Proetto}(2016)}]{Proetto_2016}%
  \BibitemOpen
  \bibfield  {author} {\bibinfo {author} {\bibfnamefont {A.}~\bibnamefont
  {Ben\'{\i}tez}}\ and\ \bibinfo {author} {\bibfnamefont {C.~R.}\ \bibnamefont
  {Proetto}},\ }\href {\doibase 10.1103/PhysRevA.94.052506} {\bibfield
  {journal} {\bibinfo  {journal} {Phys. Rev. A}\ }\textbf {\bibinfo {volume}
  {94}},\ \bibinfo {pages} {052506} (\bibinfo {year} {2016})}\BibitemShut
  {NoStop}%
\bibitem [{\citenamefont {Hollins}\ \emph {et~al.}(2017)\citenamefont
  {Hollins}, \citenamefont {Clark}, \citenamefont {Refson},\ and\ \citenamefont
  {Gidopoulos}}]{Hollins_2017}%
  \BibitemOpen
  \bibfield  {author} {\bibinfo {author} {\bibfnamefont {T.~W.}\ \bibnamefont
  {Hollins}}, \bibinfo {author} {\bibfnamefont {S.~J.}\ \bibnamefont {Clark}},
  \bibinfo {author} {\bibfnamefont {K.}~\bibnamefont {Refson}}, \ and\ \bibinfo
  {author} {\bibfnamefont {N.~I.}\ \bibnamefont {Gidopoulos}},\ }\href
  {http://stacks.iop.org/0953-8984/29/i=4/a=04LT01} {\bibfield  {journal}
  {\bibinfo  {journal} {J. Phys. Condens. Matter}\ }\textbf {\bibinfo {volume}
  {29}},\ \bibinfo {pages} {04LT01} (\bibinfo {year} {2017})}\BibitemShut
  {NoStop}%
\bibitem [{\citenamefont {Newton}(1970)}]{Newton_1970}%
  \BibitemOpen
  \bibfield  {author} {\bibinfo {author} {\bibfnamefont {R.}~\bibnamefont
  {Newton}},\ }\href {https://doi.org/10.1137/1012079} {\bibfield  {journal}
  {\bibinfo  {journal} {SIAM Review}\ }\textbf {\bibinfo {volume} {12}},\
  \bibinfo {pages} {346} (\bibinfo {year} {1970})}\BibitemShut {NoStop}%
\bibitem [{\citenamefont {Carter}(2000)}]{Carter_2000}%
  \BibitemOpen
  \bibfield  {author} {\bibinfo {author} {\bibfnamefont {A.~H.}\ \bibnamefont
  {Carter}},\ }\href {\doibase 10.1119/1.19530} {\bibfield  {journal} {\bibinfo
   {journal} {Am. J. Phys.}\ }\textbf {\bibinfo {volume} {68}},\ \bibinfo
  {pages} {698} (\bibinfo {year} {2000})}\BibitemShut {NoStop}%
\bibitem [{\citenamefont {Jayatilaka}(1998)}]{Jayatilaka_PRL.80.798}%
  \BibitemOpen
  \bibfield  {author} {\bibinfo {author} {\bibfnamefont {D.}~\bibnamefont
  {Jayatilaka}},\ }\href {\doibase 10.1103/PhysRevLett.80.798} {\bibfield
  {journal} {\bibinfo  {journal} {Phys. Rev. Lett.}\ }\textbf {\bibinfo
  {volume} {80}},\ \bibinfo {pages} {798} (\bibinfo {year} {1998})}\BibitemShut
  {NoStop}%
\bibitem [{\citenamefont {Levy}(1979)}]{Levy_1979}%
  \BibitemOpen
  \bibfield  {author} {\bibinfo {author} {\bibfnamefont {M.}~\bibnamefont
  {Levy}},\ }\href {http://www.pnas.org/content/76/12/6062.abstract} {\bibfield
   {journal} {\bibinfo  {journal} {Proc. Natl. Acad. Sci. U.S.A}\ }\textbf
  {\bibinfo {volume} {76}},\ \bibinfo {pages} {6062} (\bibinfo {year}
  {1979})}\BibitemShut {NoStop}%
\bibitem [{\citenamefont {Lieb}(1983)}]{Lieb_1983}%
  \BibitemOpen
  \bibfield  {author} {\bibinfo {author} {\bibfnamefont {E.~H.}\ \bibnamefont
  {Lieb}},\ }\href {\doibase 10.1002/qua.560240302} {\bibfield  {journal}
  {\bibinfo  {journal} {Int. J. Quantum Chem.}\ }\textbf {\bibinfo {volume}
  {24}},\ \bibinfo {pages} {243} (\bibinfo {year} {1983})}\BibitemShut
  {NoStop}%
\bibitem [{\citenamefont {Levy}\ \emph {et~al.}(1984)\citenamefont {Levy},
  \citenamefont {Perdew},\ and\ \citenamefont {Sahni}}]{LPS_1984}%
  \BibitemOpen
  \bibfield  {author} {\bibinfo {author} {\bibfnamefont {M.}~\bibnamefont
  {Levy}}, \bibinfo {author} {\bibfnamefont {J.~P.}\ \bibnamefont {Perdew}}, \
  and\ \bibinfo {author} {\bibfnamefont {V.}~\bibnamefont {Sahni}},\ }\href
  {\doibase 10.1103/PhysRevA.30.2745} {\bibfield  {journal} {\bibinfo
  {journal} {Phys. Rev. A}\ }\textbf {\bibinfo {volume} {30}},\ \bibinfo
  {pages} {2745} (\bibinfo {year} {1984})}\BibitemShut {NoStop}%
\bibitem [{\citenamefont {March}(1986)}]{MARCH_1985}%
  \BibitemOpen
  \bibfield  {author} {\bibinfo {author} {\bibfnamefont {N.}~\bibnamefont
  {March}},\ }\href {\doibase http://dx.doi.org/10.1016/0375-9601(86)90123-4}
  {\bibfield  {journal} {\bibinfo  {journal} {Phys. Lett. A}\ }\textbf
  {\bibinfo {volume} {113}},\ \bibinfo {pages} {476 } (\bibinfo {year}
  {1986})}\BibitemShut {NoStop}%
\bibitem [{\citenamefont {Levy}\ and\ \citenamefont
  {Ou-Yang}(1988)}]{LEVY_1988}%
  \BibitemOpen
  \bibfield  {author} {\bibinfo {author} {\bibfnamefont {M.}~\bibnamefont
  {Levy}}\ and\ \bibinfo {author} {\bibfnamefont {H.}~\bibnamefont {Ou-Yang}},\
  }\href {\doibase 10.1103/PhysRevA.38.625} {\bibfield  {journal} {\bibinfo
  {journal} {Phys. Rev. A}\ }\textbf {\bibinfo {volume} {38}},\ \bibinfo
  {pages} {625} (\bibinfo {year} {1988})}\BibitemShut {NoStop}%
\bibitem [{\citenamefont {Gritsenko}\ \emph {et~al.}(1994)\citenamefont
  {Gritsenko}, \citenamefont {van Leeuwen},\ and\ \citenamefont
  {Baerends}}]{Gritsenko_1994}%
  \BibitemOpen
  \bibfield  {author} {\bibinfo {author} {\bibfnamefont {O.}~\bibnamefont
  {Gritsenko}}, \bibinfo {author} {\bibfnamefont {R.}~\bibnamefont {van
  Leeuwen}}, \ and\ \bibinfo {author} {\bibfnamefont {E.~J.}\ \bibnamefont
  {Baerends}},\ }\href {\doibase 10.1063/1.468024} {\bibfield  {journal}
  {\bibinfo  {journal} {The Journal of Chemical Physics}\ }\textbf {\bibinfo
  {volume} {101}},\ \bibinfo {pages} {8955} (\bibinfo {year}
  {1994})}\BibitemShut {NoStop}%
\bibitem [{\citenamefont {Gritsenko}\ \emph {et~al.}(1998)\citenamefont
  {Gritsenko}, \citenamefont {Schipper},\ and\ \citenamefont
  {Baerends}}]{Gritsenko_1998}%
  \BibitemOpen
  \bibfield  {author} {\bibinfo {author} {\bibfnamefont {O.~V.}\ \bibnamefont
  {Gritsenko}}, \bibinfo {author} {\bibfnamefont {P.~R.~T.}\ \bibnamefont
  {Schipper}}, \ and\ \bibinfo {author} {\bibfnamefont {E.~J.}\ \bibnamefont
  {Baerends}},\ }\href {\doibase 10.1103/PhysRevA.57.3450} {\bibfield
  {journal} {\bibinfo  {journal} {Phys. Rev. A}\ }\textbf {\bibinfo {volume}
  {57}},\ \bibinfo {pages} {3450} (\bibinfo {year} {1998})}\BibitemShut
  {NoStop}%
\bibitem [{\citenamefont {Seidl}\ \emph {et~al.}(1996)\citenamefont {Seidl},
  \citenamefont {G\"orling}, \citenamefont {Vogl}, \citenamefont {Majewski},\
  and\ \citenamefont {Levy}}]{Siedl_1996}%
  \BibitemOpen
  \bibfield  {author} {\bibinfo {author} {\bibfnamefont {A.}~\bibnamefont
  {Seidl}}, \bibinfo {author} {\bibfnamefont {A.}~\bibnamefont {G\"orling}},
  \bibinfo {author} {\bibfnamefont {P.}~\bibnamefont {Vogl}}, \bibinfo {author}
  {\bibfnamefont {J.~A.}\ \bibnamefont {Majewski}}, \ and\ \bibinfo {author}
  {\bibfnamefont {M.}~\bibnamefont {Levy}},\ }\href {\doibase
  10.1103/PhysRevB.53.3764} {\bibfield  {journal} {\bibinfo  {journal} {Phys.
  Rev. B}\ }\textbf {\bibinfo {volume} {53}},\ \bibinfo {pages} {3764}
  (\bibinfo {year} {1996})}\BibitemShut {NoStop}%
\bibitem [{\citenamefont {Herman}\ and\ \citenamefont
  {Skillman}(1963)}]{Herman_PHP}%
  \BibitemOpen
  \bibfield  {author} {\bibinfo {author} {\bibfnamefont {F.}~\bibnamefont
  {Herman}}\ and\ \bibinfo {author} {\bibfnamefont {S.}~\bibnamefont
  {Skillman}},\ }\href@noop {} {\emph {\bibinfo {title} {Atomic structure
  calculations}}}\ (\bibinfo  {publisher} {Prentice-Hall Publications},\
  \bibinfo {year} {1963})\BibitemShut {NoStop}%
\bibitem [{\citenamefont {Bunge}\ \emph {et~al.}(1993)\citenamefont {Bunge},
  \citenamefont {Barrientos},\ and\ \citenamefont {Bunge}}]{Bunge_1993}%
  \BibitemOpen
  \bibfield  {author} {\bibinfo {author} {\bibfnamefont {C.}~\bibnamefont
  {Bunge}}, \bibinfo {author} {\bibfnamefont {J.}~\bibnamefont {Barrientos}}, \
  and\ \bibinfo {author} {\bibfnamefont {A.}~\bibnamefont {Bunge}},\ }\href
  {\doibase http://dx.doi.org/10.1006/adnd.1993.1003} {\bibfield  {journal}
  {\bibinfo  {journal} {Atomic Data and Nuclear Data Tables}\ }\textbf
  {\bibinfo {volume} {53}},\ \bibinfo {pages} {113 } (\bibinfo {year}
  {1993})}\BibitemShut {NoStop}%
\bibitem [{\citenamefont {Aashamar}\ \emph {et~al.}(1978)\citenamefont
  {Aashamar}, \citenamefont {Luke},\ and\ \citenamefont
  {Talman}}]{Talman_1978}%
  \BibitemOpen
  \bibfield  {author} {\bibinfo {author} {\bibfnamefont {K.}~\bibnamefont
  {Aashamar}}, \bibinfo {author} {\bibfnamefont {T.}~\bibnamefont {Luke}}, \
  and\ \bibinfo {author} {\bibfnamefont {J.}~\bibnamefont {Talman}},\ }\href
  {\doibase http://dx.doi.org/10.1016/0092-640X(78)90019-0} {\bibfield
  {journal} {\bibinfo  {journal} {Atomic Data and Nuclear Data Tables}\
  }\textbf {\bibinfo {volume} {22}},\ \bibinfo {pages} {443 } (\bibinfo {year}
  {1978})}\BibitemShut {NoStop}%
\bibitem [{\citenamefont {Kais}\ \emph {et~al.}(1993)\citenamefont {Kais},
  \citenamefont {Herschbach}, \citenamefont {Handy}, \citenamefont {Murray},\
  and\ \citenamefont {Laming}}]{Kais_1993}%
  \BibitemOpen
  \bibfield  {author} {\bibinfo {author} {\bibfnamefont {S.}~\bibnamefont
  {Kais}}, \bibinfo {author} {\bibfnamefont {D.~R.}\ \bibnamefont
  {Herschbach}}, \bibinfo {author} {\bibfnamefont {N.~C.}\ \bibnamefont
  {Handy}}, \bibinfo {author} {\bibfnamefont {C.~W.}\ \bibnamefont {Murray}}, \
  and\ \bibinfo {author} {\bibfnamefont {G.~J.}\ \bibnamefont {Laming}},\
  }\href {\doibase 10.1063/1.465765} {\bibfield  {journal} {\bibinfo  {journal}
  {J. Chem. Phys.}\ }\textbf {\bibinfo {volume} {99}},\ \bibinfo {pages} {417}
  (\bibinfo {year} {1993})}\BibitemShut {NoStop}%
\bibitem [{\citenamefont {Knight}\ \emph {et~al.}(1984)\citenamefont {Knight},
  \citenamefont {Clemenger}, \citenamefont {de~Heer}, \citenamefont {Saunders},
  \citenamefont {Chou},\ and\ \citenamefont {Cohen}}]{Knight_PRL.52.2141}%
  \BibitemOpen
  \bibfield  {author} {\bibinfo {author} {\bibfnamefont {W.~D.}\ \bibnamefont
  {Knight}}, \bibinfo {author} {\bibfnamefont {K.}~\bibnamefont {Clemenger}},
  \bibinfo {author} {\bibfnamefont {W.~A.}\ \bibnamefont {de~Heer}}, \bibinfo
  {author} {\bibfnamefont {W.~A.}\ \bibnamefont {Saunders}}, \bibinfo {author}
  {\bibfnamefont {M.~Y.}\ \bibnamefont {Chou}}, \ and\ \bibinfo {author}
  {\bibfnamefont {M.~L.}\ \bibnamefont {Cohen}},\ }\href {\doibase
  10.1103/PhysRevLett.52.2141} {\bibfield  {journal} {\bibinfo  {journal}
  {Phys. Rev. Lett.}\ }\textbf {\bibinfo {volume} {52}},\ \bibinfo {pages}
  {2141} (\bibinfo {year} {1984})}\BibitemShut {NoStop}%
\bibitem [{\citenamefont {Brack}(1993)}]{Matthias_RMP.65.677}%
  \BibitemOpen
  \bibfield  {author} {\bibinfo {author} {\bibfnamefont {M.}~\bibnamefont
  {Brack}},\ }\href {\doibase 10.1103/RevModPhys.65.677} {\bibfield  {journal}
  {\bibinfo  {journal} {Rev. Mod. Phys.}\ }\textbf {\bibinfo {volume} {65}},\
  \bibinfo {pages} {677} (\bibinfo {year} {1993})}\BibitemShut {NoStop}%
\bibitem [{\citenamefont {Harbola}\ and\ \citenamefont {Sahni}(1989)}]{MKH_89}%
  \BibitemOpen
  \bibfield  {author} {\bibinfo {author} {\bibfnamefont {M.~K.}\ \bibnamefont
  {Harbola}}\ and\ \bibinfo {author} {\bibfnamefont {V.}~\bibnamefont
  {Sahni}},\ }\href {\doibase 10.1103/PhysRevLett.62.489} {\bibfield  {journal}
  {\bibinfo  {journal} {Phys. Rev. Lett.}\ }\textbf {\bibinfo {volume} {62}},\
  \bibinfo {pages} {489} (\bibinfo {year} {1989})}\BibitemShut {NoStop}%
\bibitem [{\citenamefont {Wagner}\ \emph {et~al.}(2013)\citenamefont {Wagner},
  \citenamefont {Stoudenmire}, \citenamefont {Burke},\ and\ \citenamefont
  {White}}]{Wagner_2013}%
  \BibitemOpen
  \bibfield  {author} {\bibinfo {author} {\bibfnamefont {L.~O.}\ \bibnamefont
  {Wagner}}, \bibinfo {author} {\bibfnamefont {E.~M.}\ \bibnamefont
  {Stoudenmire}}, \bibinfo {author} {\bibfnamefont {K.}~\bibnamefont {Burke}},
  \ and\ \bibinfo {author} {\bibfnamefont {S.~R.}\ \bibnamefont {White}},\
  }\href {\doibase 10.1103/PhysRevLett.111.093003} {\bibfield  {journal}
  {\bibinfo  {journal} {Phys. Rev. Lett.}\ }\textbf {\bibinfo {volume} {111}},\
  \bibinfo {pages} {093003} (\bibinfo {year} {2013})}\BibitemShut {NoStop}%
\bibitem [{\citenamefont {Gidopoulos}(2011)}]{Gidopoulos_2011}%
  \BibitemOpen
  \bibfield  {author} {\bibinfo {author} {\bibfnamefont {N.~I.}\ \bibnamefont
  {Gidopoulos}},\ }\href {\doibase 10.1103/PhysRevA.83.040502} {\bibfield
  {journal} {\bibinfo  {journal} {Phys. Rev. A}\ }\textbf {\bibinfo {volume}
  {83}},\ \bibinfo {pages} {040502} (\bibinfo {year} {2011})}\BibitemShut
  {NoStop}%
\bibitem [{\citenamefont {Irons}\ \emph {et~al.}(2017)\citenamefont {Irons},
  \citenamefont {Furness}, \citenamefont {Ryley}, \citenamefont {Zemen},
  \citenamefont {Helgaker},\ and\ \citenamefont {Teale}}]{Teal_2017}%
  \BibitemOpen
  \bibfield  {author} {\bibinfo {author} {\bibfnamefont {T.~J.~P.}\
  \bibnamefont {Irons}}, \bibinfo {author} {\bibfnamefont {J.~W.}\ \bibnamefont
  {Furness}}, \bibinfo {author} {\bibfnamefont {M.~S.}\ \bibnamefont {Ryley}},
  \bibinfo {author} {\bibfnamefont {J.}~\bibnamefont {Zemen}}, \bibinfo
  {author} {\bibfnamefont {T.}~\bibnamefont {Helgaker}}, \ and\ \bibinfo
  {author} {\bibfnamefont {A.~M.}\ \bibnamefont {Teale}},\ }\href {\doibase
  10.1063/1.4985883} {\bibfield  {journal} {\bibinfo  {journal} {J. Chem.
  Phys.}\ }\textbf {\bibinfo {volume} {147}},\ \bibinfo {pages} {134107}
  (\bibinfo {year} {2017})}\BibitemShut {NoStop}%
\bibitem [{\citenamefont {Gritsenko}\ \emph {et~al.}(1995)\citenamefont
  {Gritsenko}, \citenamefont {van Leeuwen},\ and\ \citenamefont
  {Baerends}}]{Vlb_1995}%
  \BibitemOpen
  \bibfield  {author} {\bibinfo {author} {\bibfnamefont {O.~V.}\ \bibnamefont
  {Gritsenko}}, \bibinfo {author} {\bibfnamefont {R.}~\bibnamefont {van
  Leeuwen}}, \ and\ \bibinfo {author} {\bibfnamefont {E.~J.}\ \bibnamefont
  {Baerends}},\ }\href {\doibase 10.1103/PhysRevA.52.1870} {\bibfield
  {journal} {\bibinfo  {journal} {Phys. Rev. A}\ }\textbf {\bibinfo {volume}
  {52}},\ \bibinfo {pages} {1870} (\bibinfo {year} {1995})}\BibitemShut
  {NoStop}%
\bibitem [{\citenamefont {Gaudoin}\ and\ \citenamefont
  {Burke}(2004)}]{Gaudoin_2004}%
  \BibitemOpen
  \bibfield  {author} {\bibinfo {author} {\bibfnamefont {R.}~\bibnamefont
  {Gaudoin}}\ and\ \bibinfo {author} {\bibfnamefont {K.}~\bibnamefont
  {Burke}},\ }\href {\doibase 10.1103/PhysRevLett.93.173001} {\bibfield
  {journal} {\bibinfo  {journal} {Phys. Rev. Lett.}\ }\textbf {\bibinfo
  {volume} {93}},\ \bibinfo {pages} {173001} (\bibinfo {year}
  {2004})}\BibitemShut {NoStop}%
\bibitem [{\citenamefont {van Leeuwen}(2003)}]{Vlb_2003}%
  \BibitemOpen
  \bibfield  {author} {\bibinfo {author} {\bibfnamefont {R.}~\bibnamefont {van
  Leeuwen}},\ }\href {\doibase http://dx.doi.org/10.1016/S0065-3276(03)43002-5}
  {\bibfield  {journal} {\bibinfo  {journal} {Adv. Quantum Chem.}\ }\textbf
  {\bibinfo {volume} {43}},\ \bibinfo {pages} {25 } (\bibinfo {year}
  {2003})}\BibitemShut {NoStop}%
\bibitem [{\citenamefont {Slater}(1951)}]{Slater_1951}%
  \BibitemOpen
  \bibfield  {author} {\bibinfo {author} {\bibfnamefont {J.~C.}\ \bibnamefont
  {Slater}},\ }\href {\doibase 10.1103/PhysRev.81.385} {\bibfield  {journal}
  {\bibinfo  {journal} {Phys. Rev.}\ }\textbf {\bibinfo {volume} {81}},\
  \bibinfo {pages} {385} (\bibinfo {year} {1951})}\BibitemShut {NoStop}%
\bibitem [{\citenamefont {Schipper}\ \emph {et~al.}(1997)\citenamefont
  {Schipper}, \citenamefont {Gritsenko},\ and\ \citenamefont
  {Baerends}}]{Schipper1997}%
  \BibitemOpen
  \bibfield  {author} {\bibinfo {author} {\bibfnamefont {P.~R.~T.}\
  \bibnamefont {Schipper}}, \bibinfo {author} {\bibfnamefont {O.~V.}\
  \bibnamefont {Gritsenko}}, \ and\ \bibinfo {author} {\bibfnamefont {E.~J.}\
  \bibnamefont {Baerends}},\ }\href {\doibase 10.1007/s002140050273} {\bibfield
   {journal} {\bibinfo  {journal} {Theoretical Chemistry Accounts}\ }\textbf
  {\bibinfo {volume} {98}},\ \bibinfo {pages} {16} (\bibinfo {year}
  {1997})}\BibitemShut {NoStop}%
\bibitem [{\citenamefont {Gidopoulos}\ and\ \citenamefont
  {Lathiotakis}(2012)}]{Gidopoulos_2012}%
  \BibitemOpen
  \bibfield  {author} {\bibinfo {author} {\bibfnamefont {N.~I.}\ \bibnamefont
  {Gidopoulos}}\ and\ \bibinfo {author} {\bibfnamefont {N.~N.}\ \bibnamefont
  {Lathiotakis}},\ }\href {\doibase 10.1103/PhysRevA.85.052508} {\bibfield
  {journal} {\bibinfo  {journal} {Phys. Rev. A}\ }\textbf {\bibinfo {volume}
  {85}},\ \bibinfo {pages} {052508} (\bibinfo {year} {2012})}\BibitemShut
  {NoStop}%
\bibitem [{\citenamefont {Cuevas-Saavedra}\ \emph {et~al.}(2015)\citenamefont
  {Cuevas-Saavedra}, \citenamefont {Ayers},\ and\ \citenamefont
  {Staroverov}}]{Viktor_2015_JCP}%
  \BibitemOpen
  \bibfield  {author} {\bibinfo {author} {\bibfnamefont {R.}~\bibnamefont
  {Cuevas-Saavedra}}, \bibinfo {author} {\bibfnamefont {P.~W.}\ \bibnamefont
  {Ayers}}, \ and\ \bibinfo {author} {\bibfnamefont {V.~N.}\ \bibnamefont
  {Staroverov}},\ }\href {\doibase 10.1063/1.4937943} {\bibfield  {journal}
  {\bibinfo  {journal} {J. Chem. Phys.}\ }\textbf {\bibinfo {volume} {143}},\
  \bibinfo {pages} {244116} (\bibinfo {year} {2015})}\BibitemShut {NoStop}%
\bibitem [{\citenamefont {Nagy}(1997)}]{Nagy_1997}%
  \BibitemOpen
  \bibfield  {author} {\bibinfo {author} {\bibfnamefont {A.}~\bibnamefont
  {Nagy}},\ }\href {\doibase 10.1103/PhysRevA.55.3465} {\bibfield  {journal}
  {\bibinfo  {journal} {Phys. Rev. A}\ }\textbf {\bibinfo {volume} {55}},\
  \bibinfo {pages} {3465} (\bibinfo {year} {1997})}\BibitemShut {NoStop}%
\bibitem [{\citenamefont {Krieger}\ \emph {et~al.}(1992)\citenamefont
  {Krieger}, \citenamefont {Li},\ and\ \citenamefont {Iafrate}}]{KLI_1992}%
  \BibitemOpen
  \bibfield  {author} {\bibinfo {author} {\bibfnamefont {J.~B.}\ \bibnamefont
  {Krieger}}, \bibinfo {author} {\bibfnamefont {Y.}~\bibnamefont {Li}}, \ and\
  \bibinfo {author} {\bibfnamefont {G.~J.}\ \bibnamefont {Iafrate}},\ }\href
  {\doibase 10.1103/PhysRevA.46.5453} {\bibfield  {journal} {\bibinfo
  {journal} {Phys. Rev. A}\ }\textbf {\bibinfo {volume} {46}},\ \bibinfo
  {pages} {5453} (\bibinfo {year} {1992})}\BibitemShut {NoStop}%
\end{thebibliography}%
\end{document}